\documentclass[journal]{IEEEtran}
\usepackage{amsmath,amsfonts}
\usepackage{graphicx,amssymb}
\usepackage{algorithmic,algorithm}
\usepackage{amsthm,dsfont,multirow}
\usepackage{subfigure,url,color,capt-of,cuted}
\usepackage[noadjust]{cite}

\theoremstyle{plain}
\newtheorem{lem}{Lemma}
\newtheorem{thm}{Theorem}

\newtheorem{prop}[]{Proposition}

\newcommand\relphantom[1]{\mathrel{\phantom{#1}}}

\DeclareMathOperator{\Tr}{Tr}
\DeclareMathOperator{\Rank}{\mathrm{Rank}}
\newcommand{\Nt}{{N_\mathrm{t}}}
\newcommand{\Nr}{{N_\mathrm{r}}}

\begin{document}
	\title{ Robust and Secure Wireless Communications via Intelligent Reflecting Surfaces}
	
	\author{
		Xianghao~Yu,~\IEEEmembership{Member,~IEEE},
		Dongfang~Xu,~\IEEEmembership{Student Member,~IEEE},
		Ying~Sun,~\IEEEmembership{Member,~IEEE},
		Derrick~Wing~Kwan~Ng,~\IEEEmembership{Senior Member,~IEEE},
		and~Robert~Schober,~\IEEEmembership{Fellow,~IEEE}
		\thanks{
			This work has been presented in part at the IEEE Global Communications Conference, Waikoloa, HI, USA, Dec. 2019 \cite{yu2019enabling}.
			
			X. Yu, D. Xu, and R. Schober are with the Institute for Digital Communications, Friedrich-Alexander-University Erlangen-N\"{u}rnberg (FAU), 91054 Erlangen,
			Germany (e-mail: \{xianghao.yu, dongfang.xu, robert.schober\}@fau.de). 
			
			Y. Sun is with the School of Industrial Engineering, Purdue University, West Lafayette, IN 47907, USA (e-mail: sun578@purdue.edu).
			
			D. W. K. Ng is with the School of Electrical Engineering and Telecommunications, University of New South Wales, Sydney, NSW 2052, Australia
			(e-mail: w.k.ng@unsw.edu.au).
		}
	}
	
	\maketitle

\begin{abstract}
In this paper, intelligent reflecting surfaces (IRSs) are employed to enhance the physical layer security in a challenging radio environment. In particular, a multi-antenna access point (AP) has to serve multiple single-antenna legitimate users, which do not have line-of-sight communication links, in the presence of multiple multi-antenna potential eavesdroppers whose channel state information (CSI) is not perfectly known. Artificial noise
(AN) is transmitted from the AP to deliberately
impair the eavesdropping channels for security provisioning.
We investigate the joint design of the beamformers and AN covariance matrix  at the AP and the phase shifters at the IRSs for maximization of the system sum-rate while limiting the maximum information leakage to the potential eavesdroppers.
To this end, we formulate a robust non-convex optimization problem taking into account  the impact of the imperfect CSI of the eavesdropping channels.
To address the non-convexity of the optimization problem, an efficient algorithm is developed by capitalizing on alternating optimization, a penalty-based approach, successive convex approximation, and semidefinite relaxation.
Simulation results show that IRSs can significantly improve the system secrecy performance compared to conventional architectures without IRS. 
Furthermore, our results unveil that, for physical layer security, uniformly distributing  the reflecting elements among multiple IRSs is preferable over deploying them at a single IRS.
\end{abstract}


\IEEEpeerreviewmaketitle

\section{Introduction}
Two fundamental properties of the wireless medium, namely broadcast and superposition, make wireless transmissions inherently vulnerable to security breaches, which has been a pivotal issue in modern wireless communication systems \cite{7081071,4599222,7762075}.
Besides conventional cryptographic encryption methods employed in the application layer, advanced signal processing techniques developed for facilitating  security in the physical layer have emerged as indispensable means to guarantee secure wireless communication in recent years \cite{7762075}. The essential premise of these techniques is to exploit the intrinsic randomness of the noise and fading characteristics of wireless communication channels to limit the amount of information that can be extracted by potential eavesdroppers. 
{\color{black}Various approaches for enhancing physical layer security of wireless networks have been proposed in the literature including cooperative relaying \cite{5876337}, artificial noise (AN)-aided beamforming \cite{8349956}, and cooperative jamming \cite{5352243}. However, these existing approaches have two main demerits. First, deploying active relays and other helpers for security provisioning incurs high hardware cost and consumes additional energy. Second, in unfavorable wireless propagation environments, it is difficult to guarantee satisfactory  secrecy performance  even if AN or jamming signals are exploited.}
However, the propagation properties of wireless channels  cannot be adaptively controlled with traditional communication technologies as would be desirable for ensuring secure communication.
To overcome these two shortcomings of existing systems, a new cost-effective and energy-efficient paradigm is needed which is capable of shaping the radio propagation environment.

Benefiting from the rapid evolution of radio frequency (RF) micro-electro-mechanical systems (MEMS), programmable and reconfigurable metasurfaces have found abundant applications in different domains in recent years \cite{cui2014coding}. 
Among various types of metasurfaces, intelligent reflecting surfaces (IRSs),  a kind of passive metasurface, are particularly appealing and have been recently exploited for wireless communication system design \cite{yu2020optimal,8910627,zhang2019capacity,pan2019,zhang2019multiple,pan2019intelligent,yu2020optimal,8855810}.
Specifically, IRSs are able to change the end-to-end signal propagation direction with low-cost \emph{passive} components, e.g., printed dipoles and phase shifters, which is a revolutionary new characteristic that has not been leveraged in any of the contemporary wireless systems. 
Furthermore, metasurfaces in the form of artificial thin films can be easily coated on existing infrastructures such as walls of buildings, which reduces the operational expenditure and complexity of deploying IRSs. 
Hence, IRSs hold great promise for many applications as they provide a cost-effective approach to control the radio propagation environment while avoiding the deployment of additional power-hungry and expensive communication devices, e.g., amplify-and-forward relays \cite{8466374}. In particular, these characteristics position IRSs as a key enabler for improving the physical layer security of wireless communication systems in an economical and energy-efficient manner \cite{di2019smart}.
However, to fully unleash the potential of IRSs for security provisioning, joint optimization of the IRSs and conventional approaches to enhance physical layer security, such as transmit beamforming  and AN injection at the access point (AP), is required.

Recently, the amalgamation of IRSs and  physical layer security has received significant attention in the literature. Initial results on IRS-assisted secure wireless systems were provided in \cite{yu2019enabling,8743496,8723525}, starting from a simple network model with only one legitimate user and one eavesdropper. As a matter of fact, one of the difficulties for algorithm design for IRS-assisted systems are the highly non-convex unit modulus constraints induced by the phase shifter implementation of IRSs. Based on   semidefinite relaxation (SDR) and Gaussian randomization methods, suboptimal solutions for both the beamformer at the AP and the phase shifts at the IRS were developed in \cite{8723525}. Majorization minimization techniques were leveraged in \cite{yu2019enabling,8743496} to tackle the non-convex unit modulus constraints, which led to a better performance  compared to the classical SDR method. 
Secure wireless systems comprising multiple legitimate users and multiple eavesdroppers were investigated in \cite{xu2019resource} and \cite{guan2019intelligent}, respectively, where manifold optimization and  SDR  were applied to handle the non-convexity of the formulated problems.
To guarantee secure communication, joint optimization of transmit beamforming and AN injection at the AP were considered in both \cite{xu2019resource} and \cite{guan2019intelligent}. The authors of \cite{8742603} studied the maximization of the minimum secrecy rate among several legitimate users in the presence of multiple eavesdroppers, where the unit modulus constraints of the phase shifters were approximated by a set of convex constraints. Although this approach considerably simplifies the algorithm design, it may lead to a significant performance loss. 
All these existing works \cite{yu2019enabling,8743496,8723525,xu2019resource,guan2019intelligent,8742603} assumed that the potential eavesdroppers were equipped with a single antenna and perfect channel state information (CSI) of the eavesdropping channels was available at the AP.
However, in practice, a worst-case assumption is needed for successful security provisioning, i.e., the eavesdroppers are expected to possess more hardware resources and computational capabilities than the legitimate users, e.g., antenna elements and  interference mitigation capabilities.
Furthermore, acquiring perfect CSI of the eavesdropping channels at the AP is challenging since potential eavesdroppers are not continuously interacting with the AP and the corresponding CSI at the AP may be outdated even if the channel is only slowly time-varying. 
Therefore, the  assumptions of single-antenna eavesdroppers and perfect eavesdropper CSI at the AP are generally overly optimistic which weakens the generality and practicality of the system models considered in \cite{yu2019enabling,8743496,8723525,xu2019resource,guan2019intelligent,8742603} and the associated resource allocation algorithm designs.
To the best of the authors' knowledge, the design of robust and secure IRS-assisted wireless  communication systems with multi-antenna eavesdroppers has not been studied in the literature, yet.

To address the aforementioned issues, this paper investigates  physical layer security provisioning for IRS-assisted wireless systems, where a multi-antenna AP transmits confidential data to multiple single-antenna legitimate users in the presence of several multi-antenna potential eavesdroppers. To help establish a favorable  propagation environment for secure communication,  IRSs implemented by programmable phase shifters are deployed. The system sum-rate is maximized by taking into account the imperfect knowledge of the CSI of the eavesdropping channels, while the maximum information leakage to the potential eavesdroppers is constrained. 
To this end, the design of the considered robust and secure IRS-assisted multiuser multiple-input single-output (MISO) system is formulated as a non-convex optimization problem. 
The highly non-convex unit modulus constraints induced by the phase shifter implementation and the infinitely many inequality constraints introduced by the imperfect CSI are the main challenges for solving the problem.
First, an efficient alternating optimization (AO) based approach is developed to tackle the non-convexity of the problem, and subsequently, the design of the transmit beamformers, AN covariance matrix, and  IRS phase shifts is handled by successive convex approximation (SCA) and SDR-based methods. 
In particular, the non-convex IRS phase shift design problem with unit modulus constraints is reformulated as an equivalent rank-constrained problem, for which an effective difference of convex (d.c.) function representation is adopted to handle the rank-one constraint.
Simulation results confirm the effectiveness of the AO based algorithm compared to various baseline methods.
Furthermore, our results reveal that by deploying IRSs, favorable channel conditions are created for the legitimate users, and therefore, the physical layer security of wireless systems can be  significantly improved. Moreover, we show that  uniformly distributing the reflecting elements across multiple IRSs is beneficial for enhancing the secrecy performance.


\emph{Notations:} In this paper, the imaginary unit of a complex number is denoted by $\jmath=\sqrt{-1}$.
Matrices and vectors are denoted by boldface capital and lower-case letters, respectively. 
The set of nonnegative integers is denoted as $\mathbb{N}=\{0,1,\cdots\}$.
$\mathbb{C}^{m\times n}$ denotes the set of all $m\times n$ complex-valued matrices; 
$\mathbb{H}^{m}$ denotes the set of all $m\times m$ Hermitian matrices; 
$\mathbf{I}_m$ is the $m$-dimensional identity matrix; $\mathbf{1}_m$ represents the $m\times1$ all-ones vector; $\mathbf{0}$ represents the all-zeros matrix.
$\mathbf{X}^*$, $\mathbf{X}^T$, and $\mathbf{X}^H$ stand for the conjugate, transpose, and conjugate transpose of matrix $\mathbf{X}$, respectively.
The $\ell_2$-norm of vector $\mathbf{x}$ is denoted as $||\mathbf{x}||$. The determinant, Frobenius norm, nuclear norm, and spectral norm of matrix $\mathbf{X}$ are represented as $\det(\mathbf{X})$, $\left\Vert\mathbf{X}\right\Vert_F$, $\left\Vert\mathbf{X}\right\Vert_*$, and $\left\Vert\mathbf{X}\right\Vert_2$, respectively.
{\color{black}$\mathrm{diag}(\mathbf{x})$ denotes a diagonal matrix whose diagonal elements are extracted from vector $\mathbf{x}$, and $\mathrm{blkdiag}(\mathbf{X}_1,\cdots ,\mathbf{X}_n)$ denotes a block diagonal matrix whose diagonal components are $\mathbf{X}_1,\cdots ,\mathbf{X}_n$;}
$\mathrm{Diag}(\mathbf{X})$ represents a vector whose elements are extracted from the diagonal elements of matrix $\mathbf{X}$.
The largest eigenvalue of matrix $\mathbf{X}$ and the corresponding eigenvector are denoted by $\lambda_{\max}(\mathbf{X})$ and $\boldsymbol{\lambda}_{\max}(\mathbf{X})$, respectively.
$\triangleq$  and $\sim$ mean ``defined as'' and ``distributed as'', respectively.
$\Tr(\mathbf{X})$ and $\Rank(\mathbf{X})$ denote the trace and rank of matrix $\mathbf{X}$;
$\mathbf{X}\succeq\mathbf{0}$ indicates that $\mathbf{X}$ is a positive semidefinite (PSD) matrix.
The distribution of a circularly symmetric complex Gaussian (CSCG) vector with mean vector $\mathbf{x}$ and covariance matrix $\mathbf{\Sigma}$ is denoted
by $\mathcal{CN}(\mathbf{x},\mathbf{\Sigma})$.  For a real-valued
continuous function $f(\mathbf{X})$, $\nabla_\mathbf{X}f$ represents the gradient
of $f$ with respect to matrix $\mathbf{X}$. $\mathbb{E}[\cdot]$ represents statistical
expectation and $\left[x\right]^+=\max\{0,x\}$. The optimal value of an optimization variable $\mathbf{X}$ is denoted by $\mathbf{X}^\star$.

\section{System Model}
In this section, we first present the system model for secure IRS-assisted multiuser MISO downlink   wireless communication. Then, we discuss our assumptions regarding  CSI knowledge for system design. 
\subsection{IRS-Assisted  System Model}\label{IIA}
\begin{figure}[t]
	\centering\includegraphics[width=8.5cm]{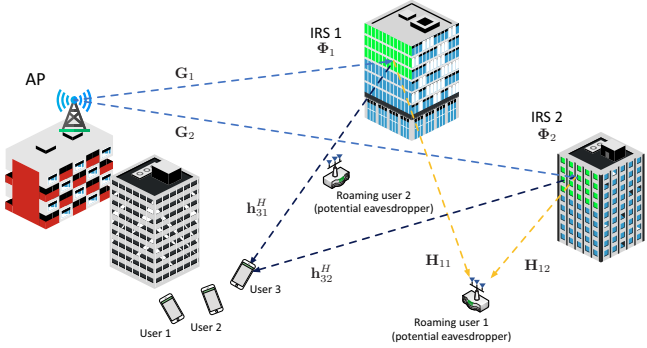}
	\caption{An IRS-assisted multiuser MISO secure communication system comprising $K=3$  legitimate users, $J=2$ roaming users, and $L=2$ IRSs coated on two buildings. The reflecting elements are represented by  green rectangles. For the ease of illustration, only the parameters of the  channels from the IRSs to potential eavesdropper $1$ and legitimate user $3$ are shown.
	}\label{model}
\end{figure}
We consider the downlink of an IRS-assisted secure communication system that comprises one AP, $K$ single-antenna legitimate users, $J$ roaming users, and $L$ IRSs, as shown in Fig. \ref{model}. The AP is   equipped with $\Nt>1$ transmit antennas.
Different from the local legitimate users, the $J$ roaming users are traveling wireless devices belonging to other communication systems and are equipped with $\Nr>1$ antennas. The roaming users may attempt to deliberately eavesdrop the  signals transmitted to the legitimate users. Therefore, we treat the $J$ roaming users as potential eavesdroppers.
In addition, $L$ passive IRSs are deployed in the network to improve the physical layer security, where IRS $l$ is equipped  with $M_l$ programmable phase shifters. 
The direct links from the AP to the legitimate users are assumed to be blocked by obstacles (e.g., buildings).
As a result, the AP communicates with the legitimate users via the IRSs and focuses its transmit beams on the IRSs. We assume that both the AP and the IRSs are deployed at sufficiently high altitudes such that the potential eavesdroppers cannot intercept the links between the AP and the IRSs.
Furthermore, as the  beams reflected by the IRSs to the legitimate users and potential eavesdroppers are typically tilting down to the ground to serve the legitimate users,  reflections between IRSs are negligible.
Therefore, the downlink received baseband signals at legitimate user $k$ and potential eavesdropper $j$ can be expressed as\footnote{For multiple IRSs,  the delays between the propagation paths reflected by different IRSs are typically much shorter than the symbol duration. For example, in a small cell network with $200$ m cell radius, the maximum delay  is $1.3$ $\mu$s while the symbol duration in the Long-Term Evolution (LTE)  standard is $70$ $\mu$s \cite{arunabha2010fundamentals}. 
Thus, intersymbol interference is not considered in \eqref{eq1} and \eqref{eq2}.
}
{\allowdisplaybreaks
\begin{align}
y_k&=\sum_{l\in\mathcal{L}}\mathbf{h}_{kl}^H\mathbf{\Phi}_l\mathbf{G}_l\left(\sum_{i\in\mathcal{K}}\mathbf{w}_is_i+\mathbf{z}\right)+n_k\notag\\
&=\mathbf{h}_k^H\mathbf{\Phi}\mathbf{G}\left(\sum_{i\in\mathcal{K}}\mathbf{w}_is_i+\mathbf{z}\right)+n_k,\quad\forall k\in\mathcal{K},\label{eq1}\\
\mathbf{y}_j&=\sum_{l\in\mathcal{L}}\mathbf{H}_{jl}\mathbf{\Phi}_l\mathbf{G}_l\left(\sum_{i\in\mathcal{K}}\mathbf{w}_is_i+\mathbf{z}\right)+\mathbf{n}_j\notag\\
&=\mathbf{H}_j\mathbf{\Phi}\mathbf{G}\left(\sum_{i\in\mathcal{K}}\mathbf{w}_is_i+\mathbf{z}\right)+\mathbf{n}_j,\quad\forall j\in\mathcal{J},\label{eq2}
\end{align}}
respectively, where $\mathcal{L}=\{1,\cdots,L\}$, $\mathcal{K}=\{1,\cdots,K\}$, $\mathcal{J}=\{1,\cdots,J\}$, $\mathbf{h}_k^H=\left[\mathbf{h}_{k1}^H,\cdots,\mathbf{h}_{kL}^H\right]$, $\mathbf{H}_j=\left[\mathbf{H}_{j1},\cdots,\mathbf{H}_{jL}\right]$, $\mathbf{\Phi}=\mathrm{blkdiag}\left(\mathbf{\Phi}_1,\cdots,\mathbf{\Phi}_L\right)$, and $\mathbf{G}=\left[\mathbf{G}_1,\cdots,\mathbf{G}_L\right]^T$.
Here,  the information-bearing signal transmitted by the AP to legitimate user $i$ and the corresponding beamforming vector are denoted as $s_i\in\mathbb{C}$ and $\mathbf{w}_i\in\mathbb{C}^{\Nt\times1}$, respectively, where $\mathbb{E}[|s_i|^2]=1$, $\forall i$, without loss of generality. 
The channel matrix from the AP to IRS $l$ is denoted as $\mathbf{G}_l\in\mathbb{C}^{M_l\times\Nt}$, while the channel vector between legitimate user $k$ and  IRS $l$  is represented by $\mathbf{h}_{kl}^H\in\mathbb{C}^{1\times M_l}$. $\mathbf{H}_{jl}\in\mathbb{C}^{\Nr\times M_l}$ is the channel matrix between eavesdropper $j$ and IRS $l$. 
The phase shift matrix $\mathbf{\Phi}_l$ at IRS $l$ is given by $\mathbf{\Phi}_l=\mathrm{diag}(\mathbf{v}_l)$,
where $\mathbf{v}_l=\left[e^{\jmath \theta_{l1}},e^{\jmath \theta_{l2}},\cdots,e^{\jmath \theta_{lM_l}}\right]^T$ and $\theta_{li}$ is the controllable phase shift introduced by the $i$-th reflecting element of IRS $l$ \cite{8811733}.
In addition, $n_k\sim\mathcal{CN}\left(0,\sigma_{\mathrm{l},k}^2\right)$ and $\mathbf{n}_j\sim\mathcal{CN}\left(\mathbf{0},\sigma_{\mathrm{e},j}^2\mathbf{I}_\Nr\right)$  represent the additive white Gaussian noise (AWGN) at legitimate user $k$ and potential eavesdropper $j$, with $\sigma_{\mathrm{l},k}^2$ and $\sigma_{\mathrm{e},j}^2$ being the corresponding noise variances, respectively.
$\mathbf{z}\in\mathbb{C}^{\Nt\times1}$ is the AN vector generated by the AP to deliberately combat the eavesdroppers. Specifically, $\mathbf{z}$ is modeled as a CSCG vector with $\mathbf{z}\sim\mathcal{CN}\left(\mathbf{\mathbf{0}},\mathbf{Z}\right)$, where $\mathbf{Z}\in\mathbb{H}^\Nt$, $\mathbf{Z}\succeq\mathbf{0}$, is the covariance matrix of the AN vector. The AN is assumed to be unknown to both the legitimate users and the potential eavesdroppers.
For notational convenience, in the rest of this paper, we drop  index $l$ in the elements of the  phase shift matrices of all IRSs and use $\mathbf{\Phi}=\mathrm{diag}\left(\mathbf{v}\right)$, where $\mathbf{v}=\left[
\mathbf{v}_1^T,\mathbf{v}_2^T,\cdots,\mathbf{v}_L^T\right]^T\triangleq\left[e^{\jmath \theta_{1}},e^{\jmath \theta_{2}},\cdots,e^{\jmath \theta_{M}}\right]^T$ and $M=\sum_{l\in\mathcal{L}}M_l$.

\newcounter{TempEqCnt}                        	
\setcounter{TempEqCnt}{\value{equation}} 		
\setcounter{equation}{3}                 		
\begin{figure*}[t]
	\begin{equation}\label{eq4}
	R_k=\log_2\left(1+\frac{\left|\mathbf{h}_k^H\mathbf{\Phi}\mathbf{G}\mathbf{w}_k\right|^2}
	{\Tr\left(\mathbf{G}^H\mathbf{\Phi}^H\mathbf{h}_k\mathbf{h}_k^H\mathbf{\Phi}\mathbf{G}\mathbf{Z}\right)+\sigma_{\mathrm{l},k}^2+\sum_{i\in\mathcal{K}\backslash \{k\}}\left|\mathbf{h}_k^H\mathbf{\Phi}\mathbf{G}\mathbf{w}_i\right|^2}\right)
	\end{equation}
	\hrule
\end{figure*}
\setcounter{equation}{\value{TempEqCnt}}

\emph{Remark 1:} The legitimate users are expected to have
limited capabilities  compared to the potential eavesdroppers. In order
to ensure secure communication under such  unfavorable conditions, we consider the worst case where all legitimate users are equipped with a single antenna while all potential eavesdroppers have multiple antennas.


{
	\color{black}\emph{Remark 2:} We assume that resource allocation is performed separately for blocked and non-blocked legitimate users, respectively, via user scheduling. 
	In particular, the AP can directly serve the non-blocked legitimate users using the direct channels without the help of IRSs.
	In this paper, we focus on the resource allocation for the blocked legitimate users which are supported by multiple IRSs.
	In IRS-assisted systems, the multi-antenna AP focuses its transmit beams on the IRSs to take full advantage of the reflecting elements.
	Moreover, we assume that the AP and IRSs are installed well above ground on top of high buildings. Thus, the legitimate users and potential eavesdroppers, which are located on the ground, cannot intercept the AP-IRS link. Therefore, the signal leakage from the AP to the legitimate users and potential eavesdroppers via the respective direct links is negligible in the considered IRS-assisted system.
}

\subsection{Channel State Information (CSI)}
{\color{black}The legitimate users transmit pilot signals to the AP via the IRSs to facilitate channel estimation. 
As a result, the AP is able to periodically acquire the CSI of the legitimate users. Several channel estimation  techniques for IRS-assisted systems have been proposed recently, e.g., \cite{de2020parafac,chen2019channel}}.
Hence, we assume the availability of perfect CSI for the AP-IRS-legitimate user links, $\mathbf{G}$ and $\{\mathbf{h}_k\}_{k=1}^K$, during the whole transmission period. 
On the other hand, for the potential eavesdroppers (roaming users), they send signals to their dedicated wireless systems rather than the AP. {\color{black}Furthermore, as the potential eavesdroppers usually try to hide their existence from the AP, they are not expected to cooperate with the AP for CSI acquisition. Therefore, although the signal	leakage from the potential eavesdroppers to the AP can still be utilized for channel estimation \cite{7448937}, the acquired CSI is expected to be coarse and outdated.}
To account for this effect, we adopt a deterministic model \cite{6781609,zheng2008robust,4838902,xu2020resource} to characterize the CSI uncertainty. Specifically, the CSI of the links between the IRSs and eavesdropper $j$ is modeled as follows
\begin{equation}\label{uncertainty}
\begin{split}
\mathbf{H}_j&=\bar{\mathbf{H}}_j+\boldsymbol{\Delta}\mathbf{H}_j,\\
\Omega_j&\triangleq\left\{\boldsymbol{\Delta}\mathbf{H}_j\in\mathbb{C}^{\Nr\times M}:||\boldsymbol{\Delta}\mathbf{H}_j||_F\le\epsilon_j\right\},\quad j\in\mathcal{J},
\end{split}
\end{equation}
where $\bar{\mathbf{H}}_j\in\mathbb{C}^{\Nr\times M}$ is the  estimate of the channel of potential eavesdropper $j$  at the AP at the beginning of the scheduling slot. The CSI estimation error for eavesdropper $j$ is modeled by the term $\boldsymbol{\Delta}\mathbf{H}_j$. Continuous set $\Omega_j$ contains all  possible  CSI estimation errors with their norms bounded by $\epsilon_j>0$. 
{\color{black}The adopted deterministic model in \eqref{uncertainty} is a flexible and general CSI uncertainty model, which is able to account for the bounded CSI uncertainty caused by different elements of the CSI acquisition process in IRS-assisted systems, e.g., quantization errors of the phase shifts at the IRSs \cite{you2019intelligent}, noisy estimation \cite{wang2019channel}, and outdated feedback.
The uncertainty radius $\epsilon_j$ represents the level of uncertainty and can be chosen
 smaller for more accurate quantization and channel estimation algorithms, and longer coherence times.
}

\section{Resource Allocation Problem Formulation}
In this section, we define the adopted performance metrics and formulate the resource allocation design as an optimization problem.
\subsection{Achievable Rate and Secrecy Rate}
According to the signal model in \eqref{eq1}, the achievable rate (bits/s/Hz) of  legitimate user $k$ is given by \eqref{eq4}, as shown on the top of this page {\color{black}\cite{1558439}}.
Then, for security provisioning, we make a  worst-case assumption regarding the capabilities of the potential eavesdroppers. In particular, we assume that the potential eavesdroppers have unlimited computational resources and therefore are able to cancel all multiuser interference before decoding the information transmitted to a given legitimate user.
Therefore, the channel capacity (bits/s/Hz) between the AP and potential eavesdropper $j$ for decoding the signal of  legitimate user $k$ is given by
\setcounter{equation}{4} 
\begin{equation}\label{Qex}
C_{j,k}=\log_2\det\left(\mathbf{I}_\Nr+\mathbf{Q}_j^{-1}\mathbf{H}_j\mathbf{\Phi}\mathbf{G}\mathbf{w}_k\mathbf{w}_k^H\mathbf{G}^H\mathbf{\Phi}^H\mathbf{H}_j^H\right),
\end{equation}
where
$
\mathbf{Q}_j=\mathbf{H}_j\mathbf{\Phi}\mathbf{G}\mathbf{Z}\mathbf{G}^H\mathbf{\Phi}^H\mathbf{H}_j^H+\sigma^2_{\mathrm{e},j}\mathbf{I}_{N_\mathrm{r}}$ is the noise covariance matrix at potential eavesdropper $j$. Hence, the maximum achievable secrecy rate between the AP and  legitimate user $k$ is  given by
\begin{equation}\label{seceq}
R_{\mathrm{s},k}=\left[R_k-\underset{ j\in\mathcal{J}}{\max}\left\{C_{j,k}\right\}\right]^+,
\end{equation}
 and  the system sum secrecy rate is  given by $R_\mathrm{s}=\sum_{k\in\mathcal{K}}R_{\mathrm{s},k}$.

\subsection{Optimization Problem Formulation}
In this paper, we adopt a worst-case robust sum-rate maximization (WCR-SRM) problem formulation
for norm-bounded CSI uncertainties \cite{6482662}. In particular,
our goal is to maximize the system sum-rate while keeping the maximum information leakage to the eavesdroppers below a desired level\footnote{Note that the considered WCR-SRM formulation with information leakage constraints offers a higher flexibility for resource allocation than the direct maximization of the secrecy rate with respect to the heterogeneous secrecy requirement of different applications, e.g., video streaming, emails, Internet-of-Things (IoT), etc. In particular, the secrecy performance of each potential eavesdropper can be controlled via the information leakage parameter $\tau_{k,j}$, which allows the system operator to strike a balance between the system sum-rate and the system secrecy rate.}. 
The joint design of the transmit beamformers and AN at the AP and the phase shifts at the IRSs can be formulated as follows
\begin{equation}\label{problem}
\begin{aligned}
&\underset{\mathbf{w}_k,\mathbf{\Phi},\mathbf{Z}\in\mathbb{H}^\Nt}{\mathrm{maximize}} && 
\sum_{k\in\mathcal{K}}R_k\\
&\mathrm{\,\,subject\thinspace to}&&\mbox{C1:}\,\sum_{k\in\mathcal{K}}\left\Vert\mathbf{w}_k\right\Vert^2+\Tr\left(\mathbf{Z}\right)\le P,\\
&&&\mbox{C2:}\,\mathbf{Z}\succeq\mathbf{0},\quad\mbox{C3:}\,\mathbf{\Phi}=\mathrm{diag}\left(\mathbf{v}
\right),\\
&&&\mbox{C4:}\,\underset{\boldsymbol{\Delta}\mathbf{H}_j\in\Omega_j}{\max}C_{j,k}\le\tau_{k,j},\quad\forall k,j.
\end{aligned}
\end{equation}
In  constraint $\mbox{C1}$, $P$ is a nonnegative parameter denoting the maximum transmit power of the AP. 
Constraint $\mbox{C2}$ and $\mathbf{Z}\in\mathbb{H}^\Nt$ are imposed as the covariance matrix of the AN is a Hermitian PSD matrix.
Since the IRSs are assumed to be implemented by a total of $M$ passive phase shifters, the equivalent  phase shift matrix $\mathbf{\Phi}$ is constrained to be a diagonal matrix with $M$ unit modulus elements, as specified by constraint $\mbox{C3}$. 
The physical layer security of the system is guaranteed by constraint $\mbox{C4}$ such that the system secrecy rate is bounded from below by $R_{\mathrm{s}}\ge\sum_{k\in\mathcal{K}}\left[R_k-\underset{ j\in\mathcal{J}}{\max}\{\tau_{k,j}\}\right]^+$. Here, $\tau_{k,j}$ is a predefined parameter which limits the maximum tolerable information leakage to potential eavesdropper $j$  for wiretapping the signal transmitted to legitimate user $k$.

{\color{black}\emph{Remark 3:}  
	Unlike existing works on IRS-assisted secure communication focusing on more favorable scenarios, e.g., \cite{yu2019enabling,8743496,8723525,xu2019resource,guan2019intelligent,8742603},  in this paper, we consider the case where the potential eavesdroppers are equipped with multiple antennas but the legitimate users have only one antenna. 
	This corresponds to a worst-case assumption where the potential eavesdroppers possess more capabilities than the legitimate users. As a result, the formulated problem \eqref{problem} is a generalization of the optimization problems considered in existing works.
	The objective function is not jointly concave with respect to $\mathbf{w}_k$, $\mathbf{\Phi}$, and $\mathbf{Z}$ because  $\mathbf{\Phi}$ is coupled with $\mathbf{w}_k$ and $\mathbf{Z}$.
	Constraint $\mbox{C3}$ is highly non-convex as the phase of each phase shifter is constrained to have unit magnitude.
	Furthermore, the CSI estimation error is taken into consideration, which leads to   infinitely many non-convex constraints in $\mbox{C4}$, which forms another challenge for solving problem \eqref{problem}.
	To the best of the authors' knowledge, this is the first work that takes into account CSI uncertainty for the design of IRS-assisted wireless systems. Although there is no general approach to solving problem \eqref{problem} optimally, in the next section, we shall propose an effective algorithm to find a locally optimal solution for problem \eqref{problem}.
	}


\section{Algorithm Design for Secure IRS-Assisted Wireless Communication}
In this section, we focus on solving the formulated optimization problem. First, we tackle the coupling of the  optimization variables via AO. In fact,
AO is a widely applicable and empirically efficient approach for handling  optimization problems involving  coupled optimization variables. It has been successfully applied to several wireless communication design problems such as hybrid precoding \cite{7397861}, resource allocation \cite{7122227}, and IRS-enabled wireless communication \cite{yu2019enabling}.
For the problem at hand, based on the principle of AO, we alternately solve for $\left\{\mathbf{w}_k,\mathbf{Z}\right\}$ and $\mathbf{\Phi}$ while fixing the other variables. This yields a stationary point solution of problem \eqref{problem} as will be detailed in the following two subsections.

\subsection{Optimization of Transmit Beamforming and AN}
We first present the optimization of the beamforming vectors $\mathbf{w}_k$ and AN covariance matrix $\mathbf{Z}$ for a given phase shift matrix $\mathbf{\Phi}$. In particular, we first convert the infinitely many constraints in constraint $\mbox{C4}$ to a finite number of equivalent constraints. To develop efficient  algorithms for optimizing $\mathbf{w}_k$ and $\mathbf{Z}$,  SDR and  SCA techniques are then leveraged to tackle the non-convexity of both the objective function and constraint $\mbox{C4}$ of problem \eqref{problem}. 
Now, for a given phase shift matrix $\mathbf{\Phi}$, the  optimization problem for the  design of $\mathbf{w}_k$ and $\mathbf{Z}$ is given by
\begin{equation}\label{subproblem1}
\begin{aligned}
&\underset{\mathbf{w}_k,\mathbf{Z}\in\mathbb{H}^\Nt}{\mathrm{maximize}} && 
\sum_{k\in\mathcal{K}}R_k\\
&\mathrm{subject\thinspace to}&&\mbox{C1:}\,\sum_{k\in\mathcal{K}}\left\Vert\mathbf{w}_k\right\Vert^2+\Tr\left(\mathbf{Z}\right)\le P,\\
&&&\mbox{C2:}\,\mathbf{Z}\succeq\mathbf{0},\\
&&&\mbox{C4:}\,\underset{\boldsymbol{\Delta}\mathbf{H}_j\in\Omega_j}{\max}C_{j,k}\le\tau_{k,j},\quad\forall k,j.\\
\end{aligned}
\end{equation}
Constraint $\mbox{C4}$ is a non-convex constraint involving infinitely many inequality constraints due to the continuity of the CSI uncertainty sets $\Omega_j$. To overcome this difficulty, we first convert the infinite number of constraints in  $\mbox{C4}$ to an equivalent form with only a finite number of constraints by leveraging the following proposition and lemma.
\begin{prop}\label{prop1}
	Constraint $\emph{\mbox{C4}}$ has the following equivalent representation
	\begin{align}
	\emph{\mbox{C4}}\Leftrightarrow\underset{\boldsymbol{\Delta}\mathbf{H}_j\in\Omega_j}{\max}&\mathbf{H}_j\mathbf{\Phi}\mathbf{G}\left[\left(2^{\tau_{k,j}}-1\right)\mathbf{Z}-\mathbf{w}_k\mathbf{w}_k^H\right]\mathbf{G}^H\mathbf{\Phi}^H\mathbf{H}_j^H\notag\\
	&+\sigma^2_{\mathrm{e},j}\left(2^{\tau_{k,j}}-1\right)\mathbf{I}_{N_\mathrm{r}}\succeq\mathbf{0}.\label{C21}
	\end{align}
\end{prop}
\begin{IEEEproof}
See Appendix \ref{appA}.
\end{IEEEproof}
While constraint $\mbox{C4}$ is transformed in Proposition \ref{prop1} into a more tractable form in terms of  linear matrix inequalities (LMIs), there are still an infinite number of such LMIs. To tackle this challenge, we resort to the following lemma to convert constraint $\mbox{C4}$ into an equivalent form involving a finite number of LMIs.

\setcounter{TempEqCnt}{\value{equation}} 		
\begin{figure*}
	\setcounter{equation}{15} 
	\begin{equation}\label{Rtk}
	\tilde{R}_k=\log_2\left(1+\frac{\Tr\left(\mathbf{M}_k\mathbf{W}_k\right)}
	{\Tr\left(\mathbf{M}_k\mathbf{Z}\right)+\sigma_{\mathrm{l},k}^2+\sum_{i\in\mathcal{K}\backslash \{k\}}\Tr\left(\mathbf{M}_k\mathbf{W}_i\right)}\right)
	\end{equation}
	\setcounter{equation}{18}
	\begin{equation}\label{gradients}
	\begin{split}
	\nabla_{\mathbf{Z}}D_1\left(\mathbf{W},\mathbf{Z}\right)&=-\frac{1}{\ln2}\sum_{j\in\mathcal{K}}
	\frac{\mathbf{M}_j}{\Tr\left(\mathbf{Z}\mathbf{M}_j\right)+\sigma_{\mathrm{l},j}^2+\sum_{i\in\mathcal{K}\backslash \{j\}}\Tr\left(\mathbf{W}_i\mathbf{M}_j\right)},\\
	\nabla_{\mathbf{W}_k}D_1\left(\mathbf{W},\mathbf{Z}\right)&=-\frac{1}{\ln2}\sum_{j\ne k}
	\frac{\mathbf{M}_j}{\Tr\left(\mathbf{Z}\mathbf{M}_j\right)+\sigma_{\mathrm{l},j}^2+\sum_{i\in\mathcal{K}\backslash \{j\}}\Tr\left(\mathbf{W}_i\mathbf{M}_j\right)}
	\end{split}
	\end{equation}
	\hrule
\end{figure*}
\setcounter{equation}{\value{TempEqCnt}}

\begin{lem}\label{lem1}
(Generalized S-Procedure \cite[Prop. 3.4]{luo2004multivariate}) Consider the quadratic matrix inequality (QMI) 
\begin{equation}
\begin{split}
&h(\mathbf{X})=\mathbf{X}^H\mathbf{A}\mathbf{X}+\mathbf{X}^H\mathbf{B}+\mathbf{B}^H\mathbf{X}+\mathbf{C}\succeq\mathbf{0},\\
&\quad\quad\quad\quad\forall \mathbf{X}\in\left\{\mathbf{Y}\mid\Tr\left(\mathbf{D}\mathbf{YY}^H\right)\le1,\mathbf{D}\succeq\mathbf{0}\right\},
\end{split}
\end{equation}
where $\mathbf{A},\mathbf{D}\in\mathbb{H}^{m}$, $\mathbf{X}, \mathbf{B}\in\mathbb{C}^{m\times n}$, and $\mathbf{C}\in\mathbb{H}^n$. 
This QMI holds if and only if there exist $p\ge0$ such that
\begin{equation}
\begin{bmatrix}
\mathbf{C}&\mathbf{B}^H\\
\mathbf{B}&\mathbf{A}
\end{bmatrix}
-p
{\begin{bmatrix}
	\mathbf{I}_n&\mathbf{0}\\
	\mathbf{0}&-\mathbf{D}
	\end{bmatrix}} \succeq\mathbf{0},
\end{equation}
provided that there exists a point $\hat{\mathbf{X}}$ such that $h(\hat{\mathbf{X}})\succeq\mathbf{0}$.
\end{lem}
Substituting \eqref{uncertainty} into \eqref{C21}, we can recast  constraint $\mbox{C4}$ as follows
\begin{equation}\label{C22}
\begin{split}
&\boldsymbol{\Delta}\mathbf{H}_j\mathbf{\Phi}\mathbf{R}_k\mathbf{\Phi}^H\boldsymbol{\Delta}\mathbf{H}_j^H+\bar{\mathbf{H}}_j\mathbf{\Phi}\mathbf{R}_k\mathbf{\Phi}^H\boldsymbol{\Delta}\mathbf{H}_j^H\\
&+\boldsymbol{\Delta}\mathbf{H}_j\mathbf{\Phi}\mathbf{R}_k\mathbf{\Phi}^H\bar{\mathbf{H}}_j^H
+\bar{\mathbf{H}}_j\mathbf{\Phi}\mathbf{R}_k\mathbf{\Phi}^H\bar{\mathbf{H}}^H_j\\
&+\sigma^2_{\mathrm{e},j}\left(2^{\tau_{k,j}}-1\right)\mathbf{I}_{N_\mathrm{r}}\succeq\mathbf{0},\\
&\quad\quad\quad\quad\quad\forall
\boldsymbol{\Delta}\mathbf{H}_j\in\left\{\mathbf{Y}\mid\Tr\left(\epsilon_j^{-2}\mathbf{YY}^H\right)\le1\right\},
\end{split}
\end{equation}
where $\mathbf{R}_k\triangleq\mathbf{G}\left[\left(2^{\tau_{k,j}}-1\right)\mathbf{Z}-\mathbf{w}_k\mathbf{w}_k^H\right]\mathbf{G}^H$. Then, by applying Lemma \ref{lem1},  constraint $\mbox{C4}$ is further expressed as
\begin{equation}
\mbox{C4}\Leftrightarrow\overline{\mbox{C4}}\mbox{:}\,\mathbf{P}_{k,j}+\mathbf{S}_j\mathbf{\Phi}\mathbf{R}_k\mathbf{\Phi}^H\mathbf{S}_j^H\succeq\mathbf{0},\quad\forall k,j,
\end{equation}
where $\mathbf{S}_j=\begin{bmatrix}
\bar{\mathbf{H}}_j^T&\mathbf{I}_M
\end{bmatrix}^T$,
\begin{equation}
\mathbf{P}_{k,j}=\begin{bmatrix}
\left[\sigma^2_{\mathrm{e},j}\left(2^{\tau_{k,j}}-1\right)-p_{k,j}\right]\mathbf{I}_{N_\mathrm{r}}&\mathbf{0}\\
\mathbf{0}&p_{k,j}\epsilon_j^{-2}\mathbf{I}_M
\end{bmatrix},
\end{equation}
and $p_{k,j}\ge0$, $\forall k,j$.
The transformed constraint $\overline{\mbox{C4}}$ involves only  $KJ$ LMI constraints, which is more amenable  for algorithm design compared to the  infinitely many constraints  in the original constraint $\mbox{C4}$. However, the resulting optimization problem is still not jointly convex with respect to $p_{k,j}$, $\mathbf{w}_k$, and $\mathbf{Z}$. To proceed, we recast the optimization problem as a rank-constrained semidefinite programming (SDP) problem. With  $\mathbf{W}_k\triangleq\mathbf{w}_k\mathbf{w}_k^H$ problem \eqref{subproblem1} can be rewritten as
\begin{align}
&\underset{p_{k,j}\ge0,\mathbf{W}_k,\mathbf{Z}\in\mathbb{H}^\Nt}{\mathrm{minimize}} && 
-\sum_{k\in\mathcal{K}}\tilde{R}_k\notag\\
&\mathrm{\quad\,\, subject\thinspace to}&&\mbox{C1:}\,\sum_{k\in\mathcal{K}}\Tr\left(\mathbf{W}_k\right)+\Tr\left(\mathbf{Z}\right)\le P,\notag\\
&&&\mbox{C2:}\,\mathbf{Z}\succeq\mathbf{0},\quad\mbox{C5:}\,\mathbf{\mathbf{W}}_k\succeq\mathbf{0},\label{eq15}\\
&&&\overline{\mbox{C4}}\mbox{:}\,\mathbf{S}_j\mathbf{\Phi}\mathbf{R}_k\mathbf{\Phi}^H\mathbf{S}_j^H+\mathbf{P}_{k,j}\succeq\mathbf{0},\quad\forall k,j,\notag\\
&&&\mbox{C6:}\,\Rank\left(\mathbf{\mathbf{W}}_k\right)\le1, \quad\forall k,\notag
\end{align}
where $\tilde{R}_k$ is shown  in \eqref{Rtk} on the top of this page and
 $\mathbf{M}_k=\mathbf{G}^H\mathbf{\Phi}^H\mathbf{h}_k\mathbf{h}_k^H\mathbf{\Phi}\mathbf{G}$. The constraints $\mathbf{W}_k\succeq\mathbf{0}$, $\mathbf{W}_k\in\mathbb{H}^\Nt$, and $\Rank\left(\mathbf{W}_k\right)\le1$ are imposed to guarantee that $\mathbf{W}_k=\mathbf{w}_k\mathbf{w}_k^H$ still holds after optimizing $\mathbf{W}_k$. All constraints of problem \eqref{eq15} are convex except for constraint $\mbox{C6}$. 
Next, we tackle the non-convexity of the objective function in problem \eqref{eq15}. To  enable efficient algorithm design, we first rewrite the objective function in form of d.c. functions, i.e.,
$-\sum_{k\in\mathcal{K}}\tilde{R}_k=N_1-D_1$,
where
\setcounter{equation}{16} 
\begin{equation}\label{eq17}
\begin{split}
N_1&=-\sum_{k\in\mathcal{K}}\log_2\left(\Tr\left(\mathbf{Z}\mathbf{M}_k\right)+\sigma_{\mathrm{l},k}^2+\sum_{i\in\mathcal{K} }\Tr\left(\mathbf{W}_i\mathbf{M}_k\right)\right),\\
D_1&=-\sum_{k\in\mathcal{K}}\log_2\left(\Tr\left(\mathbf{Z}\mathbf{M}_k\right)+\sigma_{\mathrm{l},k}^2+\sum_{i\in\mathcal{K}\backslash \{k\}}\Tr\left(\mathbf{W}_i\mathbf{M}_k\right)\right)
\end{split}
\end{equation}
are two functions that are both jointly convex in terms of $p_{k,j}$, $\mathbf{W}_k$, and $\mathbf{Z}$. Now, we adopt the SCA method to obtain a convex upper bound for the objective function in an iterative manner. To facilitate SCA, we construct a global
underestimator for function $D_1$, where we use superscript ${(t)}$ to denote the iteration index of the optimization variables. In particular, 
for any feasible point $\mathbf{W}^{(t)}\triangleq\left\{\mathbf{W}_k^{(t)}\right\}_{k=1}^K$ and $\mathbf{Z}^{(t)}$, a lower bound of function $D_1$ is given by its  first-order Taylor approximation, which can be expressed as
\begin{align}
&\relphantom{=}D_1\left(\mathbf{W},\mathbf{Z}\right)\notag\\
&\ge D_1\left(\mathbf{W}^{(t)},\mathbf{Z}^{(t)}\right)+\Tr\left(\nabla^H_{\mathbf{Z}}D_1\left(\mathbf{W}^{(t)},\mathbf{Z}^{(t)}\right)\left(\mathbf{Z}-\mathbf{Z}^{(t)}\right)\right)\notag\\
&+\sum_{k\in\mathcal{K}}\Tr\left(\nabla^H_{\mathbf{W}_k}D_1\left(\mathbf{W}^{(t)},\mathbf{Z}^{(t)}\right)\left(\mathbf{W}_k-\mathbf{W}_k^{(t)}\right)\right),
\end{align}
where the gradients of  function $D_1$ with respect to $\mathbf{Z}$ and $\mathbf{W}_k$ are given in \eqref{gradients} on the top of this page. By employing this upper bound on the objective function, the only remaining non-convexity of problem \eqref{eq15} is due to rank constraint $\mbox{C6}$. Generally, solving such a rank-constrained problem is known to be NP-hard \cite{5447068}. To overcome this issue, we adopt the SDR technique and drop the rank constraint $\mbox{C6}$.  Therefore, the resulting problem that needs to be solved at feasible point $\mathbf{W}^{(t)}$ and $\mathbf{Z}^{(t)}$ is given by
\begin{align}\setcounter{equation}{19} 
&\underset{p_{k,j}\ge0,\mathbf{W}_k,\mathbf{Z}\in\mathbb{H}^\Nt}{\mathrm{minimize}} && 
N_1
-\Tr\left(\nabla^H_{\mathbf{Z}}D_1\left(\mathbf{W}^{(t)},\mathbf{Z}^{(t)}\right)\mathbf{Z}\right)\notag\\
&&&-\sum_{k\in\mathcal{K}}\Tr\left(\nabla^H_{\mathbf{W}_k}D_1\left(\mathbf{W}^{(t)},\mathbf{Z}^{(t)}\right)\mathbf{W}_k\right)\nonumber\\
&\mathrm{\quad\,\, subject\thinspace to}&&\mbox{C1:}\,\sum_{k\in\mathcal{K}}\Tr\left(\mathbf{W}_k\right)+\Tr\left(\mathbf{Z}\right)\le P,\label{eq20}\\
&&&\mbox{C2:}\,\mathbf{Z}\succeq\mathbf{0},\quad\mbox{C5:}\,\mathbf{\mathbf{W}}_k\succeq\mathbf{0}, \quad\forall k,\nonumber\\
&&&\overline{\mbox{C4}}\mbox{:}\,\mathbf{P}_{k,j}+\mathbf{S}_j\mathbf{\Phi}\mathbf{R}_k\mathbf{\Phi}^H\mathbf{S}_j^H\succeq\mathbf{0},\quad\forall k,j.\nonumber
\end{align}
This relaxed problem is jointly convex with respect to $p_{k,j}$, $\mathbf{W}_k$, and $\mathbf{Z}$, and hence can be efficiently solved by standard convex program solvers such as CVX \cite{grant2008cvx}. However, in general, there is no guarantee that  the solution obtained by SDR satisfies the rank constraint. Nevertheless, we prove the tightness of the SDR method for problem \eqref{eq20} in the following theorem.
\setcounter{TempEqCnt}{\value{equation}} 
\begin{figure*}
\setcounter{equation}{23}
\begin{equation}\label{longeq23}
\hat{R}_k=\log_2\left(1+\frac{\Tr\left(\mathbf{L}_k\mathbf{W}_k\mathbf{L}_k^H\mathbf{V}^T\right)}
{\Tr\left(\mathbf{L}_k\mathbf{Z}\mathbf{L}_k^H\mathbf{V}^T\right)+\sigma_{\mathrm{l},k}^2+\sum_{i\in\mathcal{K}\backslash \{k\}}\Tr\left(\mathbf{L}_k\mathbf{W}_i\mathbf{L}_k^H\mathbf{V}^T\right)}\right)
\end{equation}
\hrule
\end{figure*}
\setcounter{equation}{\value{TempEqCnt}}

\begin{thm}\label{th1}
	Suppose the optimal solution of problem \eqref{eq20} is denoted as $\left\{\mathbf{W}^\star_k,\mathbf{Z}^\star,p_{k,j}^\star\right\}$, where $\Rank\left(\mathbf{W}_k^\star\right)>1$. Then, there always exists another optimal solution of problem \eqref{eq20}, denoted as
	$\left\{\tilde{\mathbf{W}}^\star_k,\tilde{\mathbf{Z}}^\star,\tilde{p}_{k,j}^\star\right\}$, which not only achieves the same objective value as $\left\{\mathbf{W}^\star_k,\mathbf{Z}^\star,p_{k,j}^\star\right\}$,  but also meets the rank constraint, i.e., $\Rank\left(\tilde{\mathbf{W}}_k^\star\right)\le1$.
	
\end{thm}
\begin{IEEEproof}
	See Appendix \ref{appB} for the proof of Theorem \ref{th1}. The construction of the optimal rank-one solution is given in \eqref{construct}.
\end{IEEEproof}
Theorem \ref{th1} indicates that we can always obtain or construct a rank-constrained optimal solution for problem \eqref{eq20}. In addition, according to Theorem \ref{th1}, the optimal beamforming vector $\mathbf{w}_k^\star$ can always be recovered from $\tilde{\mathbf{W}}_k^\star$ given in \eqref{construct} by performing Cholesky decomposition, i.e.,  $\tilde{\mathbf{W}}_k^\star=\mathbf{w}_k^\star\left(\mathbf{w}_k^\star\right)^H$.


\subsection{Optimization of Phase Shifts at IRSs}
Next, we present the optimization of  phase shift matrix $\mathbf{\Phi}$ for given  transmit beamforming vectors $\mathbf{w}_k$ and AN covariance matrix $\mathbf{Z}$. By applying Lemma \ref{lem1}, the optimization problem for the IRSs  is given by
\begin{equation}\label{eq21}\hspace{-0.3em}
\begin{aligned}
&\underset{p_{k,j}\ge0,\mathbf{\Phi}}{\mathrm{minimize}} && -
\sum_{k\in\mathcal{K}}R_k\\
&\mathrm{ subject\thinspace to}&&\mbox{C3:}\,\mathbf{\Phi}=\mathrm{diag}\left(\mathbf{v}\right),\\
&&&\overline{\mbox{C4}}\mbox{:}\,\mathbf{P}_{k,j}+\mathbf{S}_j\mathbf{\Phi}\mathbf{R}_k\mathbf{\Phi}^H\mathbf{S}_j^H\succeq\mathbf{0},\quad\forall k,j.
\end{aligned}
\end{equation}
The main difficulty in solving problem \eqref{eq21} is the unit modulus constraint $\mbox{C3}$ on the main diagonal elements of $\mathbf{\Phi}$. To the best of the authors' knowledge, there is no general approach to solve  unit modulus constrained non-convex optimization problems optimally. In this paper, we propose to rewrite  problem \eqref{eq21} by adopting $\mathbf{v}=\mathrm{Diag}\left(\mathbf{\Phi}\right)$ as the optimization variable instead of $\mathbf{\Phi}$ itself, which  paves the way for leveraging SDP and SCA methods to facilitate the design of an efficient algorithm for secure IRS-assisted systems. To start with, we first take  constraint $\overline{\mbox{C4}}$ as an example to illustrate the proposed reformulation. By performing the singular value decomposition (SVD) of $\mathbf{R}_k=\sum_{i}\mathbf{p}_{k,i}\mathbf{q}_{k,i}^H$, we have
\begin{align}
	\mathbf{S}_j\mathbf{\Phi}\mathbf{R}_k\mathbf{\Phi}^H\mathbf{S}_j^H&=\sum_i\mathbf{S}_j\mathbf{\Phi}\mathbf{p}_{k,i}\mathbf{q}_{k,i}^H\mathbf{\Phi}^H\mathbf{S}_j^H\label{eq22}\\
	&=
	\sum_i\mathbf{S}_j\mathrm{diag}\left(\mathbf{p}_{k,i}\right)\mathbf{vv}^H\mathrm{diag}\left(\mathbf{q}_{k,i}^H\right)\mathbf{S}_j^H.\notag
\end{align}

Then, the objective function can  be rewritten as a function of $\mathbf{v}$ in a similar manner as $\eqref{eq22}$.
To facilitate SDP, we define $\mathbf{V}\triangleq\mathbf{vv}^H$, and therefore problem \eqref{eq21} can be recast as
 \begin{equation}\label{eq23}
 \begin{aligned}
 &\underset{p_{k,j}\ge0,\mathbf{V}\in\mathbb{H}^M}{\mathrm{minimize}} && 
 -\sum_{k\in\mathcal{K}}\hat{R}_k\\
 &\mathrm{\,\,\, subject\thinspace to}&&\overline{\mbox{C4}}\mbox{:}\,\mathbf{P}_{k,j}+\sum_{i}\mathbf{S}_j\mathrm{diag}\left(\mathbf{p}_{k,i}\right)\mathbf{V}\\
 &&&\relphantom{\overline{\mbox{C4}}}\times\mathrm{diag}\left(\mathbf{q}_{k,i}^H\right)\mathbf{S}_j^H \succeq\mathbf{0},\quad\forall k,j,\\
 &&&\mbox{C7:}\,\mathrm{Diag}\left(\mathbf{V}\right)=\mathbf{1}_M,\quad\mbox{C8:}\,\mathbf{V}\succeq\mathbf{0},\\
 &&&\mbox{C9:}\,\mathrm{Rank}\left(\mathbf{V}\right)=1,
 \end{aligned}
 \end{equation}
 where $\hat{R}_k$ is given in \eqref{longeq23} on the top of this page and  $\mathbf{L}_k=\mathrm{diag}\left(\mathbf{h}_k^H\right)\mathbf{G}$. Constraints $\mathbf{V}\in\mathbb{H}^M$, $\mbox{C8}$, and $\mbox{C9}$ are imposed to guarantee that $\mathbf{V}=\mathbf{vv}^H$ holds after optimization. More importantly,  constraint $\mbox{C7}$ is introduced to guarantee the unit modulus constraint when recovering $\mathbf{v}$ from $\mathbf{V}$.

Recall that for the optimization of $\mathbf{W}_k$ and $\mathbf{Z}$, we have temporarily removed the rank constraint on $\mathbf{W}_k$ and proved the tightness of the SDR method. However, due to the presence of  constraint $\mbox{C7}$ induced by the unit modulus constraint,  the rank of the  solution obtained from problem \eqref{eq23} is generally larger than one \cite{5447068}. As a result, instead of applying the SDR method adopted in Section IV-A, we  handle the rank-one constraint via a different approach.
In particular, we first rewrite the rank-one constraint $\mbox{C9}$ in an equivalent form:
\setcounter{equation}{24}
\begin{equation}\label{force}
\mbox{C9}\Leftrightarrow\overline{\mbox{C9}}\mbox{:}\,\left\Vert\mathbf{V}\right\Vert_*-\left\Vert\mathbf{V}\right\Vert_2\le0.
\end{equation}
Note that for any $\mathbf{X}\in\mathbb{H}^{m\times n}$, the inequality $\left\Vert\mathbf{X}\right\Vert_*=\sum_i{\sigma_i}\ge\left\Vert\mathbf{X}\right\Vert_2=\underset{i}{\max}\{\sigma_i\}$ holds, where $\sigma_i$ is the $i$-th singular value of $\mathbf{X}$. Equality holds if and only if $\mathbf{X}$ has unit rank. 
Constraint $\overline{\mbox{C9}}$ ensures this equality, which is equivalent to enforcing constraint $\mbox{C9}$.

Constraint $\overline{\mbox{C9}}$ is in a d.c. form and therefore  still non-convex with respect to $\mathbf{V}$. To overcome the non-convexity, we apply a penalty-based method as
in \cite[Ch. 17]{nocedal2006numerical}, \cite{jiang2019over} by moving constraint $\overline{\mbox{C9}}$ into the objective function, which results in the following optimization problem
 \begin{equation}\label{penalize}
\begin{aligned}
&\underset{p_{k,j}\ge0,\mathbf{V}\in\mathbb{H}^M}{\mathrm{minimize}} && 
N_2-D_2+\frac{1}{2\rho}\left(\left\Vert\mathbf{V}\right\Vert_*-\left\Vert\mathbf{V}\right\Vert_2\right)\\
 &\mathrm{\,\,\, subject\thinspace to}&&\overline{\mbox{C4}}\mbox{:}\,\mathbf{P}_{k,j}+\sum_{i}\mathbf{S}_j\mathrm{diag}\left(\mathbf{p}_{k,i}\right)\mathbf{V}\\
&&&\relphantom{\overline{\mbox{C4}}}\times\mathrm{diag}\left(\mathbf{q}_{k,i}^H\right)\mathbf{S}_j^H \succeq\mathbf{0},\quad\forall k,j,\\
&&&\mbox{C7:}\,\mathrm{Diag}\left(\mathbf{V}\right)=\mathbf{1}_M,\quad\mbox{C8:}\,\mathbf{V}\succeq\mathbf{0},
\end{aligned}
\end{equation}
where 
\begin{equation}
\begin{split}
N_2=&-\sum_{k\in\mathcal{K}}\log_2\Bigg(\Tr\left(\mathbf{L}_k\mathbf{Z}\mathbf{L}_k^H\mathbf{V}^T\right)+\sigma_{\mathrm{l},k}^2\\
&+\sum_{i\in\mathcal{K}}\Tr\left(\mathbf{L}_k\mathbf{W}_i\mathbf{L}_k^H\mathbf{V}^T\right)\Bigg),\\
D_2=&-\sum_{k\in\mathcal{K}}\log_2\Bigg(\Tr\left(\mathbf{L}_k\mathbf{Z}\mathbf{L}_k^H\mathbf{V}^T\right)+\sigma_{\mathrm{l},k}^2\\
&+\sum_{i\in\mathcal{K}\backslash \{k\}}\Tr\left(\mathbf{L}_k\mathbf{W}_i\mathbf{L}_k^H\mathbf{V}^T\right)\Bigg),
\end{split}
\end{equation}
and $\rho>0$ is a penalty factor  penalizing the violation of constraint $\overline{\mbox{C9}}$, and the following proposition states the equivalence of problems \eqref{eq23} and \eqref{penalize}.
\setcounter{TempEqCnt}{\value{equation}} 
\begin{figure*}
	\setcounter{equation}{28}
	\begin{equation}\label{longeq28}
	\nabla_{\mathbf{V}}\tilde{D}_2=\frac{1}{2\rho}\boldsymbol{\lambda}_{\max}\left(\mathbf{V}\right)\boldsymbol{\lambda}^H_{\max}\left(\mathbf{V}\right)-\frac{1}{\ln2}\sum_{k\in\mathcal{K}}
	\frac{\mathbf{L}_k^*\left(\mathbf{Z}^T+\sum_{i\in\mathcal{K}\backslash \{k\}}\mathbf{W}_i^T\right)\mathbf{L}_k^T}{\Tr\left(\mathbf{L}_k\mathbf{Z}\mathbf{L}_k^H\mathbf{V}^T\right)+\sigma^2_{\mathrm{l},k}+\sum_{i\in\mathcal{K}\backslash \{k\}}\Tr\left(\mathbf{L}_k\mathbf{W}_i\mathbf{L}_k^H\mathbf{V}^T\right)}
	\end{equation}\hrule
\end{figure*}
\setcounter{equation}{\value{TempEqCnt}}
\begin{prop}\label{prop2}
	Let $\mathbf{V}_s$ be the optimal solution of problem \eqref{penalize} for penalty factor $\rho_s$. When $\rho_s$ is sufficiently small, i.e., for $\rho_s\to0$, then any limit point $\bar{\mathbf{V}}$ of the sequence $\left\{\mathbf{V}_s\right\}$ is an optimal solution of problem \eqref{eq23}.
\end{prop}
\begin{IEEEproof}
	See Appendix \ref{appC}.
\end{IEEEproof}
Proposition \ref{prop2} implies that we can obtain a rank-one solution $\bar{\mathbf{V}}$ by solving problem \eqref{penalize} for a sufficiently small value of penalty factor $\rho$. In other words,
the diagonal elements of the phase shift matrix $\mathbf{\Phi}$, i.e., $\mathbf{v}$, can always be recovered from the rank-one solution $\bar{\mathbf{V}}$ by performing Cholesky decomposition of $\bar{\mathbf{V}}=\mathbf{v}\mathbf{v}^H$.
Since after the proposed series of transformations,
the constraints in problem \eqref{penalize} form a convex set,
we may adopt a similar approach as in Section IV-A to tackle the d.c. objective function and employ  SCA. In particular, for any feasible point $\mathbf{V}^{(t)}$, a lower bound for $\tilde{D}_2=D_2+\frac{1}{2\rho}\left\Vert\mathbf{V}\right\Vert_2$ is obtained as 
\begin{equation}
\tilde{D}_2\left(\mathbf{V}\right)\ge 
\tilde{D}_2\left(\mathbf{V}^{(t)}\right)
+\Tr\left(\nabla^H_{\mathbf{V}}\tilde{D}_2\left(\mathbf{V}^{(t)}\right)\left(\mathbf{V}-\mathbf{V}^{(t)}\right)\right),
\end{equation}
where $\nabla_{\mathbf{V}}\tilde{D}_2$ is given in \eqref{longeq28} on the top of this page.
Therefore, the optimization problem to be solved for a given feasible point $\mathbf{V}^{(t)}$ becomes
\setcounter{equation}{29}
\begin{align}
&\underset{p_{k,j}\ge0,\mathbf{V}\in\mathbb{H}^M}{\mathrm{minimize}} && 
\frac{1}{2\rho}\left\Vert\mathbf{V}\right\Vert_*+N_2-\Tr\left(\nabla^H_{\mathbf{V}}\tilde{D}_2\left(\mathbf{V}^{(t)}\right)\mathbf{V}\right)\notag\\
&\mathrm{\,\,\, subject\thinspace to}&&\overline{\mbox{C4}}\mbox{:}\,\mathbf{P}_{k,j}+\sum_{i}\mathbf{S}_j\mathrm{diag}\left(\mathbf{p}_{k,i}\right)\mathbf{V}\notag\\
&&&\relphantom{\overline{\mbox{C4}}}\times\mathrm{diag}\left(\mathbf{q}_{k,i}^H\right)\mathbf{S}_j^H \succeq\mathbf{0},\quad\forall k,j,\label{eq28}\\
&&&\mbox{C7:}\,\mathrm{Diag}\left(\mathbf{V}\right)=\mathbf{1}_M,\quad\mbox{C8:}\,\mathbf{V}\succeq\mathbf{0}.\notag
\end{align}
This optimization problem is jointly convex with respect to $p_{k,j}$ and $\mathbf{V}$, and hence it can be efficiently solved by standard convex program solvers such as CVX \cite{grant2008cvx}. 
\begin{algorithm}[t]
	\caption{Alternating Optimization Algorithm}
	\begin{algorithmic}[1]
		\STATE Randomly construct the initial points $\mathbf{w}_k^{(0)}$ and  $\mathbf{Z}^{(0)}$. Initialize $\mathbf{\Phi}^{(0)}$ with random phases. Set convergence tolerance $\varepsilon$ and iteration index $t=0$;
		\REPEAT 
		\STATE Find $\mathbf{W}_k^{(t+1)}$ and $\mathbf{Z}^{(t+1)}$ by solving problem \eqref{eq20} for the given $\mathbf{\Phi}=\mathbf{\Phi}^{(t)}$;
		\STATE Set ${\mathbf{V}}^{(t)}=\mathrm{Diag}\left(\mathbf{\Phi}^{(t)}\right)\mathrm{Diag}\left(\mathbf{\Phi}^{(t)}\right)^H$;
		\STATE Solve problem  \eqref{eq28} for given $\mathbf{W}_k=\mathbf{W}_k^{(t+1)}$ and $\mathbf{Z}=\mathbf{Z}^{(t+1)}$, and update ${\mathbf{V}}^{(t+1)}$;
		\STATE Decompose ${\mathbf{V}}^{(t+1)}=\mathbf{v}^{(t+1)}\left(\mathbf{v}^{(t+1)}\right)^H$ and update $\mathbf{\Phi}^{(t+1)}=\mathrm{diag}\left(\mathbf{v}^{(t+1)}\right)$;
		\STATE $t\leftarrow t+1$;
		\UNTIL $\frac{\sum_{k\in\mathcal{K}}\left(R_k^{(t)}-R_k^{(t-1)}\right)}{\sum_{k\in\mathcal{K}}R_k^{(t-1)}}\le\varepsilon$
	\end{algorithmic}
\end{algorithm}

The overall AO algorithm proposed in this section is summarized in \textbf{Algorithm 1}. 
Note that the minimum values of problems \eqref{eq20} and \eqref{eq28} serve as upper bounds for the optimal values of problems \eqref{eq15} and \eqref{penalize}, respectively.
By iteratively solving problems \eqref{eq20} and \eqref{eq28} optimally in Steps 3 and 5 in \textbf{Algorithm 1}, we can monotonically tighten these upper bounds.
In this way, the objective values achieved by the sequence $\left\{\mathbf{w}^{(t)}, \mathbf{Z}^{(t)},  \mathbf{\Phi}^{(t)}\right\}_{t\in\mathbb{N}}$
 form a non-increasing sequence that converges to a stationary value in polynomial time, and any limit point of the sequence $\left\{\mathbf{w}^{(t)}, \mathbf{Z}^{(t)},  \mathbf{\Phi}^{(t)}\right\}_{t\in\mathbb{N}}$ is a stationary point of problem \eqref{problem} \cite{7547360}.
 Furthermore, the computational complexity of each iteration of the proposed AO algorithm is given by
$\mathcal{O}\left(\log\frac{1}{\varepsilon}\left(\left(\sqrt{\Nt}+\sqrt{M}\right)K^3J^3+\left({\Nt}^{\frac{5}{2}}+M^{\frac{5}{2}}\right)K^2J^2\right)\right)$, where $\mathcal{O}(\cdot)$ is the big-O notation\footnote{\color{black}According to \cite[Th. 3.12]{polik2010interior}, the computational complexity of an SDP problem with $m$ SDP constraints, where each constraint involves an $n\times n$ PSD matrix, is given by $\mathcal{O}\left(\sqrt{n}\log\frac{1}{\varepsilon}\left(mn^3+m^2n^2+m^3\right)\right)$. For problems \eqref{eq20} and \eqref{eq28}, we have $m=KJ+1$, $n=\Nt$ and $m=KJ+1$, $n=M$, respectively.}.

\section{Simulation Results}\label{SecV}
\begin{figure}
	\centering
	\includegraphics[height=6cm]{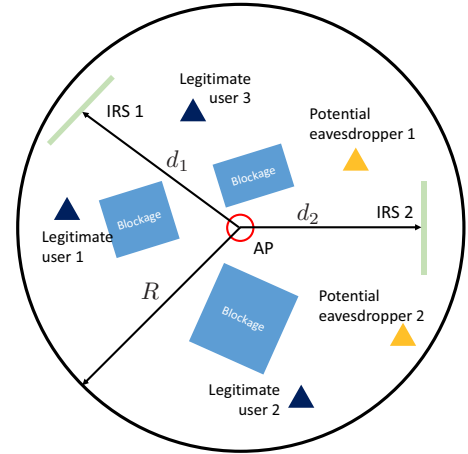}
	\captionof{figure}{Simulation setup for an IRS-assisted multiuser MISO secure communication system, which consists of $K=3$ legitimate users, $J=2$ potential eavesdroppers, and $L=2$ IRSs.}
	\label{top}
\end{figure}

\begin{table}
	\centering\caption{System parameters}
	\begin{tabular}{|l|l|}
		\hline
		Carrier center    frequency              & $2.4$ GHz                  \\ \hline
		Path loss exponents for channels with &\multirow{2}{*}{$2$ and $4$}\\
		and without LoS components, $\alpha_l$ &\\\hline
		Ricean factors for channels with &\multirow{2}{*}{$5$ and $0$}\\
		and without LoS components, $\beta_l$ &\\\hline
		Noise power at the legitimate users and&\multirow{2}{*}{$-90$ dBm}\\
		potential eavesdroppers, $\sigma_{\mathrm{l},k}^2$ and $\sigma_{\mathrm{e},j}^2$ &\\\hline
		Cell radius, $R$&$200$ m\\\hline
		Penalty factor, $\rho$&$5\times10^{-4}$\\\hline
		Convergence tolerance, $\varepsilon$&$10^{-3}$\\\hline
	\end{tabular}
\end{table}

\subsection{Simulation Setup}
The schematic system model for the simulated single-cell network is shown in Fig. \ref{top}. The AP is located at the center of the cell with radius $R$. The legitimate users and potential eavesdroppers are randomly and uniformly distributed in the cell. 
Since our motivation for deploying IRSs in secure wireless systems is to establish favorable communication links for  legitimate users that would otherwise be  blocked, we consider a scenario where $L$ IRSs are deployed to improve the secrecy performance of $K$ legitimate users whose direct links to the AP are blocked by obstacles. Each potential eavesdropper is equipped with $\Nr=2$ antennas {\color{black}unless specified otherwise}. The channel matrix $\mathbf{G}_l$ between the AP and  IRS $l$ is modeled as  follows
\begin{equation}
\mathbf{G}_l=\sqrt{L_0d_l^{-\alpha_l}}\left(\sqrt{\frac{\beta_l}{1+\beta_l}}\mathbf{G}_l^\mathrm{L}+\sqrt{\frac{1}{1+\beta_l}}\mathbf{G}_l^\mathrm{N}\right),
\end{equation}
where $L_0=\left(\frac{\lambda_{c}}{4\pi}\right)^2$ is a constant with $\lambda_{c}$ being the wavelength of the center frequency of the information
carrier. The distance between the AP and IRS $l$ is denoted by $d_l$ while $\alpha_l$ is the corresponding path loss exponent. The small-scale fading is assumed to be Ricean fading with Ricean factor $\beta_l$.
The line-of-sight (LoS) and non-LoS components are represented by $\mathbf{G}_l^\mathrm{L}$ and $\mathbf{G}_l^\mathrm{N}$,  which are modeled as the product of the array response vectors of the  transceivers and Rayleigh fading, respectively. The other channels in the system are generated in a similar manner as we expect that legitimate users with LoS links to the IRSs are served and the potential eavesdroppers may also position themselves in the LoS of the IRSs.
To facilitate the presentation, we define the maximum normalized estimation error of the eavesdropping channels as $\kappa_j={\epsilon_j}/{\left\Vert\mathbf{\bar{\mathbf{H}}}_j\right\Vert_F}$, $\forall j$. 
The system parameters used for our simulations are listed in Table I.
\subsection{Baseline Schemes}
We adopt two baseline schemes for comparison. For baseline 1, we adopt simple design choices without performing iterative optimization. In particular, we  adopt   maximum ratio transmission (MRT)  for transmit beamforming, apply an isotropic radiation pattern for AN injection, and implement the IRSs with random phases. As for  MRT, we set $\mathbf{w}_k=\sqrt{\varrho_k}\frac{\mathbf{G}^H\mathbf{\Phi}^H\mathbf{h}_k}{\left\Vert\mathbf{G}^H\mathbf{\Phi}^H\mathbf{h}_k\right\Vert}$, where $\varrho_k$ is the power allocated to legitimate user $k$, which is optimized together with the power allocated to the AN to satisfy the total transmit power and secrecy constraints as stated in \eqref{problem}. 
For baseline 2, we evaluate the system performance when IRSs are not deployed\footnote{{\color{black}If IRSs are not used, the information carrying beams emitted by the AP are typically down tilted and energy is beamformed towards the ground to reach the legitimate users, and therefore, different from the scenario described in Remark 2, the  signals received at the legitimate users and eavesdroppers via the direct links are not negligible. Therefore, the direct links have to be taken into consideration for baseline 2.}}. 
In this case, the legitimate users are blocked by infrastructures, and therefore the channels between the AP and the legitimate users are assumed to be non-LoS. As the potential eavesdroppers can move freely in the network, the channels between the AP and the eavesdroppers are assumed to be still LoS-dominated, which is an unfavorable scenario for secure communication. 
In  baseline 2, we optimize the beamforming vectors and AN covariance matrix at the AP by setting $\mathbf{\Phi}=\mathbf{I}_\Nt$ and solving problem \eqref{subproblem1}.

 In the following, we investigate the impact of the different system parameters by focusing on the case of one IRS in Sections \ref{sub1} to V-H while  multi-IRS systems are studied in  Section V-I.
 
 \begin{figure}[t]\color{black}
 	\centering
 	\includegraphics[height=6cm]{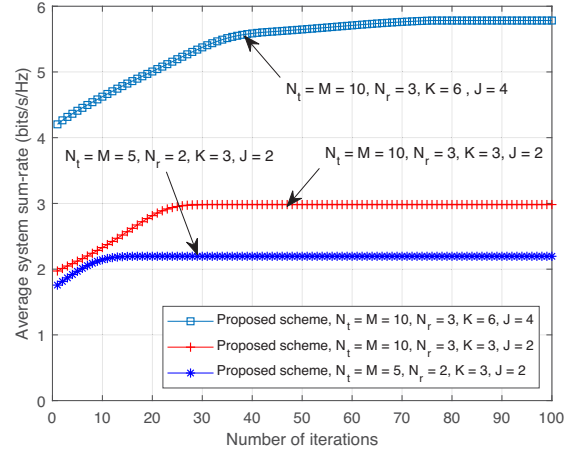}
 	\caption{Convergence of the proposed algorithm for
 		different values of $N_\mathrm{t}$, $N_\mathrm{r}$, $M$, $K$, and $J$. The system parameters are set as $P=30$ dBm,  $d_1=60$ m, $\kappa_j^2=0.1$, and $\tau_{k,j}=1$ bit/s/Hz.}
 	\label{fig0}
 \end{figure}

\subsection{Convergence of the Proposed Algorithm}\label{sub1}
In Fig. \ref{fig0}, we investigate the convergence of the proposed algorithm for different numbers of antenna elements, $\Nt$, IRS reflecting elements, $M$, and legitimate users,  $K$. 
As can be observed from Fig. \ref{fig0}, the proposed algorithm monotonically converges for all considered values of $\Nt$, $M$, and $K$.
{\color{black}In particular, for $N_\mathrm{t}=M=5$, $N_\mathrm{r}=2$, $K=3$, $J=2$, the proposed algorithm converges after around 15 iterations on average.
	For the case with more antennas and IRS reflecting elements, i.e., $N_\mathrm{t}=M=10$, $N_\mathrm{r}=3$, $K=3$, $J=2$, the proposed algorithm needs additional 15 iterations on average to converge since the dimensions of the solution space of problems \eqref{eq20} and \eqref{eq28} scale with $N_\mathrm{t}$, $N_\mathrm{r}$, and $M$.
	For the case with more legitimate users and potential eavesdroppers, i.e., $N_\mathrm{t}=M=10$, $N_\mathrm{r}=3$, $K=6$, $J=4$, the proposed algorithm needs considerably more iterations to converge since the numbers of optimization variables and constraints in problems \eqref{eq20} and \eqref{eq28} both increase with the numbers of legitimate users $K$ and potential eavesdroppers $J$.
	In summary, the number of iterations required for the proposed algorithm to converge is less sensitive to the numbers of antennas and  IRS reflecting elements than to the numbers of legitimate users and potential eavesdroppers.}
\begin{figure}[t]\color{black}
		\centering
		\includegraphics[height=6cm]{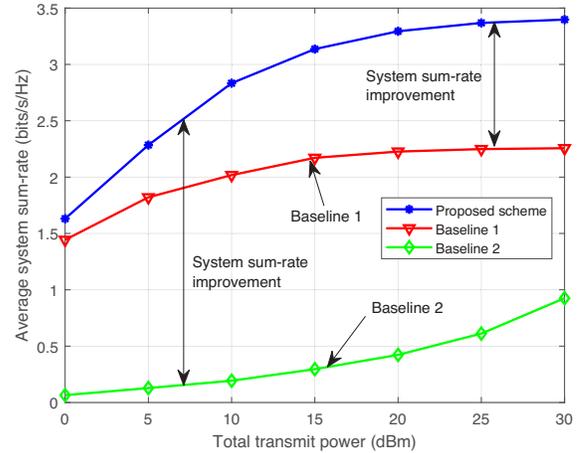}
		\caption{Average system sum-rate (bits/s/Hz) versus the total transmit power of the AP (dBm) for $N_\mathrm{t}=M=6$, $N_\mathrm{r}=3$, $K=4$, $J=2$, $d_1=115$ m, $\kappa_j^2=0.1$, and $\tau_{k,j}=1$ bit/s/Hz. The double sided arrows indicate the performance gain achieved by the proposed scheme over the baseline schemes.}
		\label{fig1}
\end{figure}

\begin{figure*}[t]
	\centering
	\subfigure[Average system secrecy rate (bits/s/Hz) versus the maximum tolerable channel capacity of the potential eavesdroppers (bits/s/Hz).]
	{
		\centering\includegraphics[height=6.2cm]{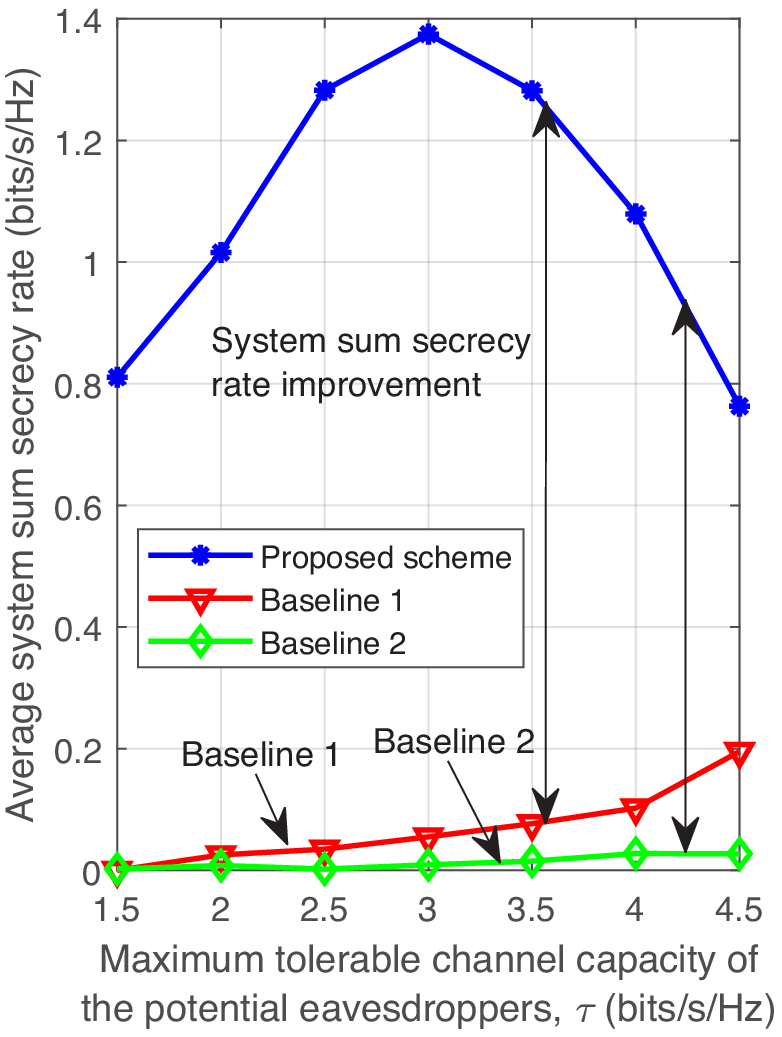}\label{fig21}
	}\quad
	\subfigure[Power allocated to beamforming and AN versus the maximum tolerable channel capacity of the potential eavesdroppers (bits/s/Hz).]
	{
		\centering\includegraphics[height=6.2cm]{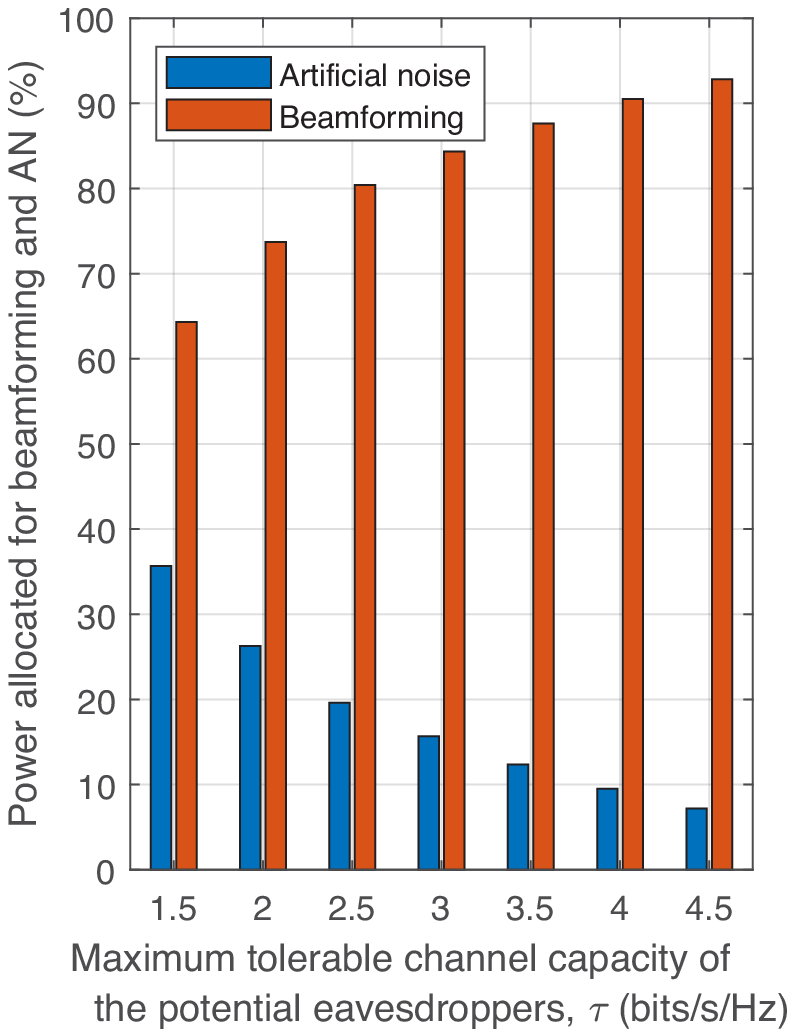}
		\label{fig22}
	}\quad
	\subfigure[Average system sum-rate (bits/s/Hz) and sum of maximum channel capacities (bits/s/Hz) of the  potential eavesdroppers for the proposed scheme.]
	{
		\centering\includegraphics[height=6.2cm]{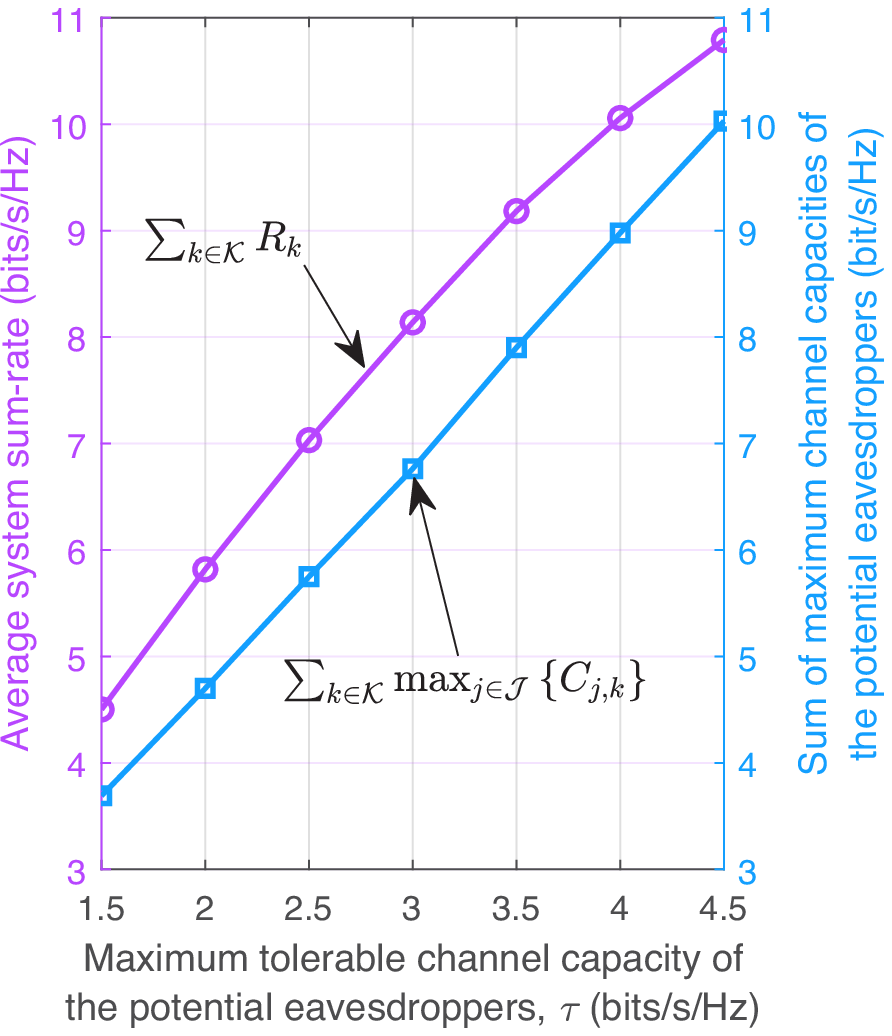}
		\label{fig23}
	}
	\caption{The system parameters are set as $\Nt=M=6$, $K=3$, $J=2$, $d_1=50$ m,  $\kappa_j^2=0.1$, and $P=10$ dBm.}
\end{figure*}

\subsection{Average System Sum Rate Versus the Maximum Transmit Power}
In Fig. \ref{fig1}, we show the average system sum-rate versus the maximum total transmit power at the AP, $P$, while limiting the information leakage to the potential eavesdroppers to $\tau_{k,j}=1$ bit/s/Hz, $\forall k,j$. As can be observed from Fig. \ref{fig1}, the system sum-rate increases monotonically with the maximum transmit power budget. This is because by applying the proposed optimization framework, the signal-to-interference-plus-noise ratios (SINRs) of the legitimate users can be improved by  providing them with additional transmit power, which leads to an improvement of the system sum-rate.
Fig. \ref{fig1} also shows that the average system sum-rate of the proposed scheme exceeds that of both considered baseline schemes by a considerable margin. 
In particular, for baseline  1,  the fixed MRT beamforming is unable to fully exploit the extra {\color{black}degrees of freedom (DoFs)} introduced by the additional transmit power. 
In fact, the  MRT strategy in baseline 1 is unable to mitigate multiuser interference resulting in performance saturation in the high transmit power regime.
For baseline  2, the performance loss compared to the proposed scheme is mainly due to the huge path loss of the AP-legitimate user links as the legitimate users are assumed to be heavily attenuated. 
Fig. \ref{fig1} reveals the huge performance gains enabled by deploying an IRS to establish an LoS propagation environment for the legitimate users. 

\subsection{Secrecy Rate Versus the Maximum Tolerable Channel Capacity of the Eavesdroppers}
Fig. \ref{fig21} depicts the average secrecy rate defined in \eqref{seceq} versus the maximum tolerable channel capacity of the potential eavesdroppers, $\tau_{k,j}$. We assume that the  maximum tolerable capacity of all eavesdroppers are identical, i.e., $\tau_{k,j}=\tau$, $\forall k,j$. 
As can be observed,  the system secrecy rate is almost zero if the IRS is not available (baseline 2), which is mainly due to the poor channel conditions between the AP and the legitimate users. In other words, for blocked legitimate users, secure wireless communication cannot be guaranteed without the  favorable propagation environment created by the IRS.
Furthermore,  the average secrecy rate achieved by the proposed scheme is significantly higher than that of  baseline  1, which confirms  the effectiveness of the proposed  algorithm. 
As shown in Fig. \ref{fig22}, when $\tau$ is small, a large portion of the transmit power is allocated to transmitting AN to deteriorate the achievable rates of the potential eavesdroppers, and therefore, there is less power left for information beamforming to maximize the system sum-rate. As $\tau$ increases, the constraints on the performance of the eavesdroppers are relaxed, and hence, more transmit power is allocated to beamforming and to improving the system sum-rate, as can be observed in  Figs. \ref{fig22} and \ref{fig23}.
Recall that the secrecy rate is given by $R_{\mathrm{s}}=\sum_{k\in\mathcal{K}}\left[R_k-\underset{ j\in\mathcal{J}}{\max}\left\{C_{j,k}\right\}\right]^+$. For large $\tau$, the maximum  channel capacity of the potential eavesdroppers $\sum_{k\in\mathcal{K}}\underset{ j\in\mathcal{J}}{\max}\left\{C_{j,k}\right\}$ grows  faster than the system sum-rate $\sum_{k\in\mathcal{K}}R_k$, as can be observed in Fig. \ref{fig23}. Therefore, the secrecy rate  decreases for large $\tau$, see also Fig. \ref{fig21}.
Hence,  if system sum secrecy rate maximization is desired, $\tau$ has to be carefully chosen. We note that the maximum secrecy rate can be found by solving the considered problem for different values of $\tau$.

\subsection{Energy Efficiency Evaluation}
\begin{figure}[t]
	\centering
	\includegraphics[height=6cm]{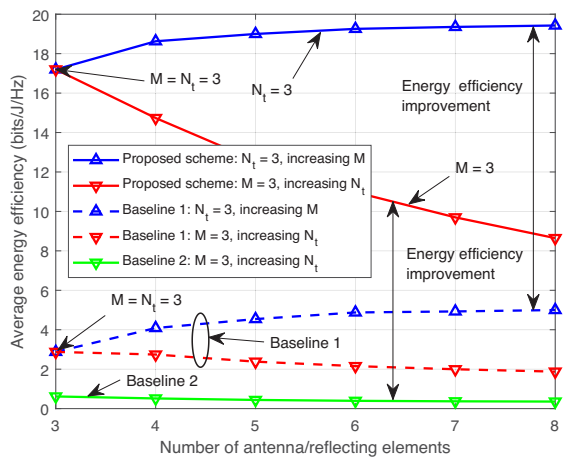}
	\captionof{figure}{Average system energy efficiency (bits/J/Hz) versus the number of antennas, $\Nt$, or reflecting elements, $M$, for $K=3$, $J=2$, $d_1=50$ m,  $\kappa_j^2=0.1$, $\tau_{k,j}=1$ bit/s/Hz, and $P=5$ dBm.}
	\label{fig3}
\end{figure}

IRSs are advocated as energy-efficient devices for assisting wireless communication.
In Fig. \ref{fig3}, we investigate the energy efficiency versus the number of antenna elements at the AP and the number of reflecting elements at the IRS. We adopt a linear power consumption model, and the energy efficiency (bits/J/Hz) is defined as the ratio of
the system sum-rate to the total power consumption of the system \cite{6529189}
\begin{equation}
\eta_\mathrm{EE}=\frac{\sum_{k\in\mathcal{K}}R_k}{\frac{1}{\mu}P+\Nt P_\mathrm{t}+P_o+P_\mathrm{I}},
\end{equation}
where $\mu$ is the power amplifier efficiency, and $P_\mathrm{t}$ accounts for the circuit power
consumption introduced by deploying one antenna element, which is mainly caused by the power consumed by the corresponding RF chain. $P_o$  is the static circuit power of the AP,  consumed by the cooling system, power supply, etc.,  and $P_\mathrm{I}$ is the  power consumption of the IRS controller. Following \cite{6529189}, we set $\mu=0.32$, $P_\mathrm{t}=35$ mW, $P_\mathrm{I}=20$ mW, and  $P_o=34$ mW.
In Fig. \ref{fig3}, we show the average energy efficiency as a function of the number of reflecting elements at the IRS for $\Nt=3$ antennas (black curves) and as a function of the number of transmit antennas for $M=3$ IRS reflecting elements (red curves). 
As can be observed from Fig. \ref{fig3}, increasing the number of IRS  elements  leads to  an improvement of the energy efficiency. The reason behind this is twofold. On the one hand, as they are passive devices, enlarging the IRS size causes little additional power consumption. On the other hand, additional phase shifters can reflect more of the 
power  received from the AP, which leads to a power
gain, and  provide more flexibility for resource allocation, which improves the beamforming gain for the links from the  IRS to the legitimate users.
In contrast, the energy efficiency is a monotone decreasing function with respect to the number of antennas at the AP. This is mainly because more power-hungry RF chains are required as the number of antennas increases, which  outweighs the system sum-rate gain introduced by deploying more antennas.
Fig. \ref{fig3} clearly demonstrates the superiority of IRS-assisted secure communication systems compared to conventional multi-antenna secure wireless systems in terms of energy efficiency.

\subsection{Average System Sum-Rate Versus the Number of Legitimate Users}
\begin{figure}[t]
	\centering
	\includegraphics[height=6cm]{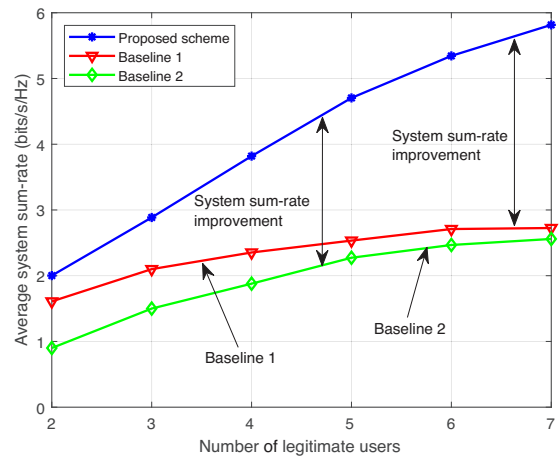}
	\captionof{figure}{Average system sum-rate (bits/s/Hz) versus the number of legitimate users, $K$, for $\Nt=M=10$,  $J=2$, $d_1=50$ m,  $P=25$ dBm,  $\kappa_j^2=0.1$,  and $\tau_{k,j}=1$ bit/s/Hz.}
	\label{fig4}
\end{figure}

Fig. \ref{fig4} shows the average system sum-rate versus the number of legitimate users, $K$. As can be observed, the system sum-rates achieved by the proposed scheme and both baseline schemes monotonically increase with the number of legitimate users. This is due to the fact that both the proposed scheme and the two baseline schemes are able to exploit multiuser diversity.
However, the system sum-rate and its growth rate are substantially lower for the baseline schemes   compared to the proposed scheme. 
In particular, the MRT strategy adopted in baseline 1 fails to mitigate  multiuser interference, which quickly leads to a saturation of  the performance.
On the other hand, without the LoS links created by the IRS for the legitimate users, baseline 2 achieves the lowest system sum-rate among the three considered schemes.
These observations underline  the superiority of  IRS-assisted secure multiuser systems and the proposed  algorithm.
\subsection{CSI Uncertainty}
\begin{figure*}[t]
	\centering
	\subfigure[Average system sum-rate (bits/s/Hz) versus the maximum normalized channel estimation error variance.]
	{
		\centering\includegraphics[height=6cm]{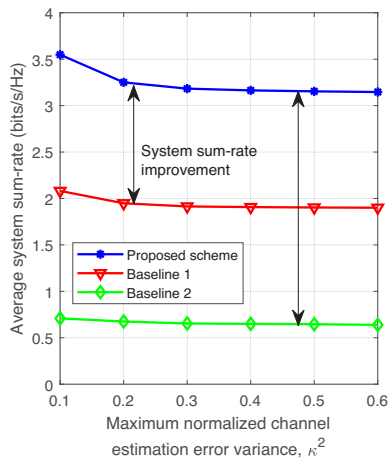}\label{fig51}
	}\quad
	\subfigure[Outage probability of potential eavesdroppers versus the target SINR (dB).]
	{
		\centering\includegraphics[height=6cm]{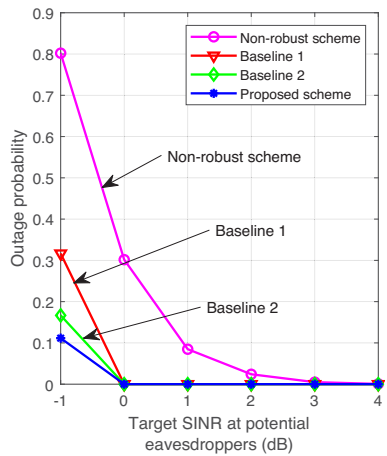}
		\label{fig52}
	}
	\caption{The system parameters are set as $\Nt=M=6$, $K=3$,  $J=2$, $d_1=20$ m,  $P=10$ dBm, and $\tau_{k,j}=1$ bit/s/Hz.}
\end{figure*}

Fig. \ref{fig51} shows the average system sum-rate versus the normalized maximum channel estimation error variance. We assume that all eavesdropping channels have the same maximum normalized estimation error variance, i.e., $\kappa_j=\kappa$, $\forall j$. 
As can be observed, for the proposed scheme and both baseline schemes, the average system sum-rates decrease as the quality of the CSI degrades. In particular, the worse the quality of the estimated CSI  is, the more difficult it is for the AP to perform accurate beamforming and efficient AN jamming, which results in a lower achievable system sum-rate. In addition, the proposed scheme significantly outperforms both baseline schemes over the entire range of considered estimation error variances, which indicates that the proposed scheme can exploit the  spatial DoFs available for security provisioning more efficiently than the two baseline schemes  even in the presence of CSI uncertainty.
In Fig. \ref{fig52}, the outage probability of the potential eavesdroppers versus the target  SINR is investigated for $\kappa^2=0.1$. The outage probability is defined as the probability that the received SINR values at the eavesdroppers are higher than a predefined target SINR. 
For  comparison, we also investigate the outage probability of a non-robust scheme. The considered non-robust scheme treats the estimated CSI of the eavesdropping channels, i.e., $\bar{\mathbf{H}}_j$, as perfect CSI.
As can be observed, a significant outage probability reduction can be achieved via robust optimization with the proposed scheme and even with the baseline schemes compared to the non-robust scheme, especially in the low target SINR regime. 
Specifically, the outage probability of the proposed scheme is $10\%$ at $-1$ dB whereas the outage probability of the non-robust scheme is as high as $80\%$.
Furthermore, as we set the maximum tolerable channel capacity of the potential eavesdroppers to $\tau_{k,j}=1$ bit/s/Hz, the outage probabilities of the proposed scheme and both baseline schemes are  zero when the target SINRs are no less than $0$ dB (as $\log_2(1+1)=1$ bit/s/Hz). In contrast, the outage probability for the non-robust scheme at $0$ dB is still higher than $30\%$.
These results  confirm  the robustness of the proposed scheme.

\subsection{Single IRS or Multiple IRSs?}
In this subsection, we compare the average system sum-rates achieved by deploying a single IRS and two IRSs in the secure wireless network shown in Fig. \ref{top}. For the scenario when two IRSs are deployed, they are symmetrically located to the East and West of the AP. We assume that there are in total $M=10$ reflecting elements available for deployment of the two IRSs. In particular, $M_1$ reflecting elements are allocated to IRS 1, and $M_2=M-M_1$ elements are assigned to IRS 2.
\begin{figure}[t]\centering
	\includegraphics[height=6cm]{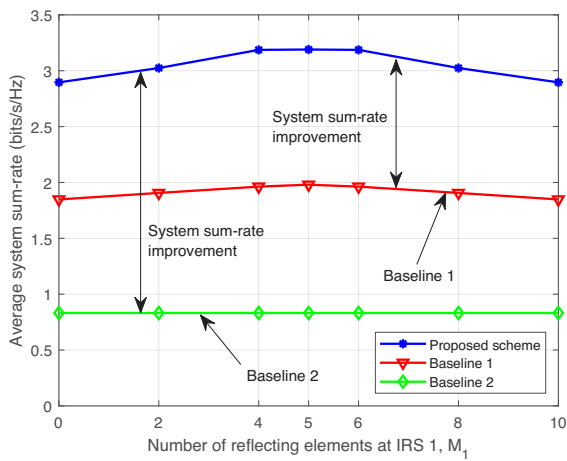}\caption{Average system sum-rate (bits/s/Hz) versus the number of reflecting elements at IRS 1, for $M=10$, $\Nt=6$,  $K=3$,  $J=2$, $d_1=d_2=20$ m,  $P=25$ dBm, $\kappa_j^2=0.1$, and $\tau_{k,j}=1$ bit/s/Hz.}
	\label{fig6}
\end{figure}
As can be observed in Fig. \ref{fig6}, the proposed scheme considerably outperforms both baseline schemes for all IRS settings. In addition, it is  beneficial to deploy two IRSs compared to only one. 
{\color{black}
	This is because multiple IRSs create multiple independent propagation paths which introduces macro diversity, and thus, facilitates the establishment of strong end-to-end LoS channels from the AP to the legitimate users. 
	Specifically, when multiple IRSs are deployed in the network, the distance between each legitimate user and its nearest IRS is reduced, which improves the system sum-rate. On the other hand, although the distance between each potential eavesdropper and its nearest IRS is also shortened, our proposed algorithm can effectively degrade the channel quality of the eavesdroppers such that the secrecy performance is improved.}
In fact, the peak system sum-rate is achieved by uniformly distributing the reflecting elements among the deployed IRSs, i.e., $5$ reflecting elements for each IRS. As indicated by Fig. \ref{fig3}, the performance gain diminishes as the number of reflecting elements  at one IRS grows large compared to the number of users. 
Therefore, a biased allocation with exceedingly many reflecting elements at one of the IRSs is not favorable.

\section{Conclusions}
In this paper, we incorporated IRSs in multiuser MISO systems to achieve secure communication  in the presence of multiple multi-antenna potential eavesdroppers for the challenging case where the legitimate users do not have a LoS link to the AP. 
The transmit beamformers, AN covariance matrix, and IRS phase shifts  were jointly optimized to maximize the system sum-rate while satisfying secrecy constraints for the potential eavesdroppers. 
We focused on the robust design of such IRS-assisted secure wireless systems taking into account the imperfection of the  CSI of the eavesdropping channels.
An efficient AO algorithm that yields a stationary solution of the formulated non-convex optimization problem was proposed.
In particular, the resulting challenging unit modulus constrained optimization problem was transformed to a rank constrained problem, for which an effective d.c. function representation was employed to facilitate the application of SDR and SCA techniques.
Simulation results have verified the tremendous potential of IRSs to improve the physical layer security of future wireless communication systems in an energy-efficient manner.
Our results also confirmed the robustness of the proposed scheme with respect to imperfect CSI and its ability to exploit multiuser diversity.
Furthermore, design guidelines for multi-IRS systems were provided. In particular, the uniform distribution of the reflecting elements among multiple IRSs was shown to be favorable for improving  physical layer security and exploiting macro diversity gains.

\appendix
\subsection{Proof of Proposition \ref{prop1}}\label{appA}
	According to  Sylvester's determinant identity $\det\left(\mathbf{I}+\mathbf{AB}\right)=\det\left(\mathbf{I}+\mathbf{BA}\right)$, we have
\begin{align}
&\relphantom{\Leftrightarrow}\underset{\boldsymbol{\Delta}\mathbf{H}_j\in\Omega_j}{\max}\log_2\det\big(\mathbf{I}_\Nr+\mathbf{Q}_j^{-1}\mathbf{H}_j\mathbf{\Phi}\mathbf{G}\mathbf{w}_k\nonumber\\
&\hspace{12em}\times\mathbf{w}_k^H\mathbf{G}^H\mathbf{\Phi}^H\mathbf{H}_j^H\big)\le\tau_{k,j}\nonumber\\
&\Leftrightarrow\underset{\boldsymbol{\Delta}\mathbf{H}_j\in\Omega_j}{\max}\log_2\left(1+\mathbf{w}_k^H\mathbf{G}^H\mathbf{\Phi}^H\mathbf{H}_j^H\mathbf{Q}_j^{-1}\mathbf{H}_j\mathbf{\Phi}\mathbf{G}\mathbf{w}_k\right)\le\tau_{k,j}\nonumber\\
&\Leftrightarrow\underset{\boldsymbol{\Delta}\mathbf{H}_j\in\Omega_j}{\max}\Tr\left(\mathbf{Q}_j^{-1}\mathbf{H}_j\mathbf{\Phi}\mathbf{G}\mathbf{w}_k\mathbf{w}_k^H\mathbf{G}^H\mathbf{\Phi}^H\mathbf{H}_j^H\right)\le2^{\tau_{k,j}}-1\nonumber\\
&\overset{(a)}{\Leftrightarrow}\underset{\boldsymbol{\Delta}\mathbf{H}_j\in\Omega_j}{\max}\lambda_{\max}\big(\mathbf{Q}_j^{-1/2}\mathbf{H}_j\mathbf{\Phi}\mathbf{G}\mathbf{w}_k\\
&\hspace{8em}\times\mathbf{w}_k^H\mathbf{G}^H\mathbf{\Phi}^H\mathbf{H}_j^H\mathbf{Q}_j^{-1/2}\big)\le2^{\tau_{k,j}}-1\nonumber\\
&\Leftrightarrow\underset{\boldsymbol{\Delta}\mathbf{H}_j\in\Omega_j}{\max}\left(2^{\tau_{k,j}}-1\right)\mathbf{Q}_j-\mathbf{H}_j\mathbf{\Phi}\mathbf{G}\mathbf{w}_k\mathbf{w}_k^H\mathbf{G}^H\mathbf{\Phi}^H\mathbf{H}_j^H\succeq\mathbf{0},\nonumber
\end{align}
where ${(a)}$ holds since matrix $\mathbf{Q}_j^{-1}\mathbf{H}_j\mathbf{\Phi}\mathbf{G}\mathbf{w}_k\mathbf{w}_k^H\mathbf{G}^H\mathbf{\Phi}^H\mathbf{H}_j^H$ has rank one, and therefore its maximum eigenvalue is the only  non-zero eigenvalue. By substituting the expression of $\mathbf{Q}_j$ defined after \eqref{Qex} into the last LMI, we  obtain the result in Proposition \ref{prop1}.

\subsection{Proof of Theorem \ref{th1}}\label{appB}
 We first express problem \eqref{eq20} in its epigraph form as follows\footnote{To keep the presentation concise, in this proof, we simplify the notations $\nabla_{\mathbf{Z}}D_1\left(\mathbf{W}^{(t)},\mathbf{Z}^{(t)}\right)$ and $\nabla_{\mathbf{W}_k}D_1\left(\mathbf{W}^{(t)},\mathbf{Z}^{(t)}\right)$ to  $\nabla_{\mathbf{Z}}D_1$ and $\nabla_{\mathbf{W}_k}D_1$, respectively.}
\begin{align}
&\underset{\eta_k,p_{k,j}\ge0,\mathbf{W}_k,\mathbf{Z}\in\mathbb{H}^\Nt}{\mathrm{minimize}} && 
-\sum_{k\in\mathcal{K}}\Tr\left(\mathbf{W}_k\nabla_{\mathbf{W}_k}^HD_1\right)\notag\\
&&&-\Tr\left(\mathbf{Z}\nabla_{\mathbf{Z}}^HD_1\right)-\sum_{k\in\mathcal{K}}\log_2\left(\eta_k+\sigma_{\mathrm{l},k}^2\right)\notag\\
&\mathrm{\quad\quad subject\thinspace to}&&\mbox{C1},\mbox{C2},\overline{\mbox{C4}},\mbox{C5},\label{epi}\\
&&&\mbox{C10:}\,\sum_{i\in\mathcal{K} }\Tr\left(\mathbf{W}_i\mathbf{M}_k\right)\notag\\
&&&\relphantom{\mbox{C10:}}+\Tr\left(\mathbf{Z}\mathbf{M}_k\right)\ge\eta_k,\quad\forall k.\notag
\end{align}
This problem is jointly convex with respect to the optimization variables and satisfies Slater's constraint qualification {\color{black}\cite[Ch. 5.2.3]{boyd2004convex}}. Therefore, strong duality holds and the Lagrangian function in terms of $\mathbf{W}_k$ is given by
\begin{equation}
\begin{split}
\mathcal{L}=&-\sum_{k\in\mathcal{K} }\Tr\left(\mathbf{W}_k\nabla_{\mathbf{W}_k}^HD_1\right)
+\gamma\sum_{k\in\mathcal{K} }\Tr\left(\mathbf{W}_k\right)\\
&-\sum_{k\in\mathcal{K} }\delta_{k}\sum_{i\in\mathcal{K} }\Tr\left(\mathbf{W}_i\mathbf{M}_k\right)-\sum_{k\in\mathcal{K} }\Tr\left({\mathbf{\Upsilon}_k}\mathbf{W}_k\right)\\
&+\sum_{k\in\mathcal{K}}\sum_{j\in\mathcal{J}}\Tr\left(\mathbf{\Omega}_{k,j}\mathbf{S}_j\mathbf{\Phi}\mathbf{GW}_k\mathbf{G}^H\mathbf{\Phi}^H\mathbf{S}_j^H\right)+\xi,\label{lag}
\end{split}
\end{equation}
where $\xi$ includes all terms that do not involve $\mathbf{W}_k$. The Lagrange multipliers associated with constraints $\mbox{C1}$, $\overline{\mbox{C4}}$, $\mbox{C5}$, and $\mbox{C10}$ in problem \eqref{epi} are denoted by $\gamma\ge0$, $\mathbf{\Omega}_{k,j}\in\mathbb{H}^{M+\Nr}$, $\mathbf{\Upsilon}_k\in\mathbb{H}^\Nt$, and $\delta_k\ge0$, respectively. Then, we investigate the structure of $\mathbf{W}_k$ by checking the Karush-Kuhn-Tucker (KKT) conditions for problem \eqref{epi}, which are given by
\begin{align}
&\mbox{K1:}\,\mathbf{\Upsilon}_k^\star\mathbf{W}_k^\star=\mathbf{0},\quad
\mbox{K2:}\,\gamma^\star\ge0,\delta_k^\star\ge0,\mathbf{\Omega}_{k,j}^\star\succeq\mathbf{0},\mathbf{\Upsilon}^\star_k\succeq\mathbf{0},\notag\\
&\mbox{K3:}\,\nabla_{\mathbf{W}_k}\mathcal{L}\left(\mathbf{W}_k^\star\right)=\mathbf{0},
\end{align}
where $\gamma^\star$, $\mathbf{\Omega}^\star_{k,j}$, $\mathbf{\Upsilon}^\star_k$, and $\delta_k^\star$ are the optimal Lagrangian multipliers for the dual problem of \eqref{epi}. Since constraint $\mbox{C10}$ is active at the optimal solution, we have $\delta_k>0$. 
 KKT condition $\mbox{K3}$ can be further expressed as
$
\mathbf{\Upsilon}_k^\star=
\mathbf{B}_k^\star-\delta_k^\star\mathbf{M}_k,
$
where
\begin{equation}
\begin{split}
\mathbf{B}_k^\star&=\gamma^\star\mathbf{I}_{N_\mathrm{t}}-\nabla_{\mathbf{W}_k}D_1\left(\mathbf{W}^{(t)}\right)-\sum_{i\in\mathcal{K}\backslash\{k\}}\delta_i^\star\mathbf{M}_i\\
&\relphantom{=}+\sum_{j\in\mathcal{J}}\mathbf{G}^H\mathbf{\Phi}^H\mathbf{S}_j^H\mathbf{\Omega}_{k,j}^\star\mathbf{S}_j\mathbf{\Phi G}.
\end{split}
\end{equation}

First, we discuss the case when matrix $\mathbf{B}^\star$ is full rank, i.e., $\mathrm{Rank}\left(\mathbf{B}_k^\star\right)=N_\mathrm{t}$. Recall that $\mathbf{M}_k=\mathbf{G}^H\mathbf{\Phi}^H\mathbf{h}_k\mathbf{h}_k^H\mathbf{\Phi}\mathbf{G}$ is a rank-one matrix. Then, we obtain for the rank of matrix $\mathbf{\Upsilon}_k^\star$,
\begin{equation}\label{eq27}
\Rank\left(\mathbf{\Upsilon}_k^\star\right)=\Rank\left(\mathbf{B}_k^\star-\delta_k^\star\mathbf{M}_k\right)\ge\Nt-1.
\end{equation}
The last inequality holds because $\Rank\left(\mathbf{A}-\mathbf{B}\right)\ge\Rank\left(\mathbf{A}\right)- \Rank\left(\mathbf{B}\right)$ for any $\mathbf{A}$ and $\mathbf{B}$ of the same dimension.
Suppose  $\Rank\left(\mathbf{\Upsilon}_k^\star\right)=\Nt$, then $\mbox{K1}$ implies that $\mathbf{W}_k^\star=\mathbf{0}$, i.e., $\Rank\left(\mathbf{W}_k^\star\right)=0$.
Suppose $\Rank\left(\mathbf{\Upsilon}_k^\star\right)=\Nt-1$, then $\mbox{K1}$ implies that the null space of $\mathbf{\Upsilon}_k$ is spanned by a vector, and  $\mathbf{W}_k^\star$   can be expressed as 
\begin{equation}
\mathbf{W}_k^\star=\mathbf{a}_k^\star\left(\mathbf{a}_k^\star\right)^H.
\end{equation}
Hence, the optimal solution $\mathbf{W}_k^\star$ of problem \eqref{eq20} satisfies $\mbox{C6}$   if $\mathrm{Rank}\left(\mathbf{B}_k^\star\right)=N_\mathrm{t}$.

Next, we exploit \cite[Prop. 4.1]{6728676} and study the case when matrix $\mathbf{B}_k^\star$ is not full rank, i.e.,  $r=\Rank\left(\mathbf{B}_k^\star\right)<\Nt$.
Let $\mathbf{\Pi}_k^\star\in\mathbb{C}^{\Nt\times(\Nt-r)}$ be the orthonormal basis of $\mathbf{B}_k^\star$'s null space, whose columns are $\{\boldsymbol{\pi}_{k,i}^\star\}_{i=1}^{\Nt-r}$. Since $\mathbf{\Upsilon}_k$ is the Lagrangian  multiplier associated with the PSD constraint $\mbox{C5}$, we have
\begin{align}
&\relphantom{\Rightarrow}\mathbf{\Upsilon}_k^\star\succeq\mathbf{0}\notag\\
&\Rightarrow \left(\boldsymbol{\pi}_{k,i}^\star\right)^H\mathbf{\Upsilon}_k^\star\boldsymbol{\pi}_{k,i}^\star=\left(\boldsymbol{\pi}_{k,i}^\star\right)^H\left(\mathbf{B}_k^\star-\delta_k^\star\mathbf{M}_k\right)\boldsymbol{\pi}_{k,i}^\star\ge0\notag\\
&\Rightarrow\delta_{k}^\star\left|\left(\boldsymbol{\pi}_{k,i}^\star\right)^H\mathbf{G}^H\mathbf{\Phi}^H\mathbf{h}_k\right|^2\le0.
\end{align}
Since constraint $\mbox{C10}$ is active at the optimal solution, we have $\delta_k>0$. Therefore,
\begin{equation}\label{eqortho}
\begin{split}
&\relphantom{\Rightarrow}\left|\left(\boldsymbol{\pi}_{k,i}^\star\right)^H\mathbf{G}^H\mathbf{\Phi}^H\mathbf{h}_k\right|^2=0\Rightarrow\mathbf{M}_k\mathbf{\Pi}_k^\star=\mathbf{0}\\
&\Rightarrow\mathbf{B}_k^\star\mathbf{\Pi}_k^\star-\delta_k^\star\mathbf{M}_k\mathbf{\Pi}_k^\star=\mathbf{\Upsilon}_k^\star\mathbf{\Pi}_k^\star=\mathbf{0},
\end{split}
\end{equation}
which means that $\mathbf{\Pi}_k$ spans $\Nt-r$ orthogonal dimensions of $\mathbf{\Upsilon}_k^\star$'s null space.
Therefore, let $\mathbf{N}_k^\star$ be the orthonormal basis of $\mathbf{\Upsilon}_k^\star$'s null space, where $\Rank\left(\mathbf{N}_k^\star\right)\ge\Nt-r$. 
In addition, we have 
\begin{equation}
\begin{split}
&\relphantom{\Rightarrow}\Rank\left(\mathbf{\Upsilon}_k^\star\right)
\overset{(b)}{\ge} r-1\\
&\Rightarrow
\Rank\left(\mathbf{N}^\star_k\right)=\Nt-\Rank\left(\mathbf{\Upsilon}_k^\star\right)\le\Nt-r+1,
\end{split}
\end{equation}
where $(b)$ follows from the same inequality as  \eqref{eq27}.
As a result, the rank of $\mathbf{N}^\star_k$ is either  $\Nt-r$ or $\Nt-r+1$. Suppose $\Rank\left(\mathbf{N}^\star_k\right)=\Nt-r$, according to \eqref{eqortho}, we have
$\mathbf{N}^\star_k=\mathbf{\Pi}^\star_k$. According to $\mbox{K1}$ and  \eqref{eqortho}, $\mathbf{W}_k^\star$ can be expressed as
$\mathbf{W}^\star_k=\sum_{i=1}^{\Nt-r}c_{k,i}^\star\boldsymbol{\pi}_{k,i}^\star\left(\boldsymbol{\pi}_{k,i}^\star\right)^H$, where $c_{k,i}^\star\in\mathbb{R}$ is a scaling factor for $\boldsymbol{\pi}_{k,i}^\star$. However, in this case, we have
 $\mathbf{W}_k^\star\mathbf{M}_k=\mathbf{0}$ since we have proved that $\mathbf{M}_k\mathbf{\Pi}_k^\star=\mathbf{0}$ in \eqref{eqortho}. This  means that there is no information received by  legitimate user $k$ even if we allocate power to the beamformer for  legitimate user $k$. Hence, this cannot correspond to the  optimal solution of problem \eqref{eq20}. Suppose $\Rank\left(\mathbf{N}_k^\star\right)=\Nt-r+1$, then we can express $\mathbf{N}^\star_k$ as $\mathbf{N}_k^\star=\begin{bmatrix}
 \mathbf{\Pi}_k^\star&\mathbf{a}_k^\star
 \end{bmatrix}$ and therefore the optimal solution $\mathbf{W}^\star_k$ can be written in the form of
 \begin{equation}
 \mathbf{W}_k^\star=a_k^\star\mathbf{a}^\star_k\left(\mathbf{a}^\star_k\right)^H+\sum_{i=1}^{\Nt-r}c_{k,i}^\star\boldsymbol{\pi}_{k,i}^\star\left(\boldsymbol{\pi}_{k,i}^\star\right)^H,
 \end{equation}
 where $c_{k,i}^\star>0$, $a_k^\star\in\mathbb{R}$ is a scaling factor for $\mathbf{a}^\star_k$, $\left(\mathbf{\Pi}_k^\star\right)^H\mathbf{a}_k^\star=\mathbf{0}$ since $\mathbf{N}_k^\star$ is the orthonormal basis of $\mathbf{\Upsilon}_k^\star$'s null space, and we have  $\mathbf{\Upsilon}_k^\star\mathbf{a}_k^\star=\mathbf{0}$ according to \eqref{eqortho}.
Therefore, we propose to construct an optimal rank-one $\tilde{\mathbf{W}}_k^\star$ when matrix $\mathbf{B}_k^\star$ is not full rank as follows:
\begin{equation}\label{construct}
\begin{split}
\tilde{\mathbf{\mathbf{W}}}_k^\star&=\mathbf{W}_k^\star-\sum_{i=1}^{\Nt-r}c_{k,i}^\star\boldsymbol{\pi}_{k,i}^\star\left(\boldsymbol{\pi}_{k,i}^\star\right)^H=a_k^\star\mathbf{a}^\star_k\left(\mathbf{a}^\star_k\right)^H,\\
\tilde{\mathbf{\mathbf{Z}}}^\star&=\mathbf{Z}^\star+\sum_{i=1}^{\Nt-r}c_{k,i}^\star\boldsymbol{\pi}_{k,i}^\star\left(\boldsymbol{\pi}_{k,i}^\star\right)^H,\quad \tilde{p}_{k,j}^\star={p}_{k,j}^\star.
\end{split}
\end{equation}
Note that with the construction in \eqref{construct}, we obtain a rank-one solution $\tilde{\mathbf{\mathbf{W}}}_k^\star$ and a PSD matrix $\tilde{\mathbf{\mathbf{Z}}}^\star$. Hence, the remaining tasks for proving the tightness of SDR are to check the feasibility of $\tilde{\mathbf{W}}_k^\star$ and  $\tilde{\mathbf{Z}}^\star$, and to show that this construction yields the same optimal objective value as $\mathbf{W}_k^\star$ and $\mathbf{Z}^\star$.  It is easy to see that the construction in \eqref{construct} does not affect the value of   $N$ and constraint \mbox{C1} in \eqref{eq17} due to the fact that  $\mathbf{Z}^\star+\sum_{k\in\mathcal{K}}\mathbf{W}^\star_k=\tilde{\mathbf{Z}}^\star+\sum_{k\in\mathcal{K}}\tilde{\mathbf{W}}^\star_k$.
For  constraint $\overline{\mbox{C4}}$ in problem \eqref{eq20}, we have
\begin{equation}
\begin{split}
&\relphantom{=}\mathbf{P}_{k,j}+\mathbf{S}_j\mathbf{\Phi}\mathbf{G}\left[\left(2^{\tau_{k,j}}-1\right)\tilde{\mathbf{Z}}^\star-\tilde{\mathbf{W}}_k^\star\right]\mathbf{G}^H\mathbf{\Phi}^H\mathbf{S}_j^H\\
&=\mathbf{P}_{k,j}+\mathbf{S}_j\mathbf{\Phi}\mathbf{G}\left[\left(2^{\tau_{k,j}}-1\right){\mathbf{Z}}^\star-{\mathbf{W}}_k^\star\right]\mathbf{G}^H\mathbf{\Phi}^H\mathbf{S}_j^H\\
&\relphantom{=}+\mathbf{S}_j\mathbf{\Phi}\mathbf{G}\mathbf{D}_k^\star\mathbf{G}^H\mathbf{\Phi}^H\mathbf{S}_j^H,
\end{split}
\end{equation}
where 
\begin{equation}
\begin{split}
\mathbf{D}_k^\star&=\left(2^{\tau_{k,j}}-1\right)\sum_{j\ne k}\sum_{i=1}^{\Nr-r}c_{j,i}^\star\boldsymbol{\pi}_{j,i}^\star\left(\boldsymbol{\pi}_{j,i}^\star\right)^H\\
&\relphantom{=}+2^{\tau_{k,j}}\sum_{i=1}^{\Nr-r}c_{k,i}^\star\boldsymbol{\pi}_{k,i}^\star\left(\boldsymbol{\pi}_{k,i}^\star\right)^H\succeq\mathbf{0}.
\end{split}
\end{equation} 
Therefore, the LMIs in $\overline{\mbox{C4}}$ still hold for $\tilde{\mathbf{\mathbf{W}}}_k^\star$ and  $\tilde{\mathbf{\mathbf{Z}}}^\star$.
Finally, we show that the  construction in \eqref{construct} does not change the value of the last two terms in the objective function of problem \eqref{eq20}. 
According to \eqref{gradients}, we have
\begin{equation}
\begin{split}
&\relphantom{=}
-\Tr\left(\mathbf{Z}^\star\nabla_{\mathbf{Z}}^HD_1\right)-\sum_{k\in\mathcal{K}}\Tr\left(\mathbf{W}_k^\star\nabla_{\mathbf{W}_k}^HD_1\right)\\
&=\frac{1}{\ln2}\sum_{j\in\mathcal{K}}\frac{\Tr\left[\left(\mathbf{Z}^\star+\sum_{k\ne j}\mathbf{W}_k^\star\right)\mathbf{M}_j\right]}{\Tr\left[\left(\mathbf{Z}^{(t)}+\sum_{i\ne j}\mathbf{W}_i^{(t)}\right)\mathbf{M}_j\right]+\sigma_{\mathrm{l},j}^2},
\end{split}
\end{equation}
where 
\begin{align}
&\relphantom{=}
\left(\tilde{\mathbf{Z}}^\star+\sum_{k\ne j}\tilde{\mathbf{W}}_k^\star\right)\mathbf{M}_j\notag\\
&=\left({\mathbf{Z}}^\star+\sum_{k\ne j}{\mathbf{W}}_k^\star\right)\mathbf{M}_j+\left(\sum_{i=1}^{\Nr-r}c^\star_{j,i}\boldsymbol{\pi}^\star_{j,i}\left(\boldsymbol{\pi}^\star_{j,i}\right)^H\right)\mathbf{M}_j\notag\\
&=\left({\mathbf{Z}}^\star+\sum_{k\ne j}{\mathbf{W}}_k^\star\right)\mathbf{M}_j.
\end{align}
The last equality holds since we have proved that $\mathbf{M}_k\mathbf{\Pi}_k^\star=\mathbf{0}$  in \eqref{eqortho},
which completes the proof of Theorem \ref{th1}.

\subsection{Proof of Proposition \ref{prop2}}\label{appC}
Denote the objective function of problem \eqref{eq23} as $f\left(\mathbf{V}\right)$, and let $\mathbf{V}^\star$ be its optimal solution, that is,
$
f\left(\mathbf{V}^\star\right)\le f\left(\mathbf{V}\right),
$ for all $\mathbf{V}$ that satisfy\footnote{In this proof, we  investigate the effect of  moving constraint $\overline{\mbox{C9}}$ into the objective function of problem \eqref{eq23}. Therefore, we do not consider constraints $\overline{\mbox{C4}}$, $\mbox{C7}$, and $\mbox{C8}$ as they are common constraints shared by both problems \eqref{eq23} and \eqref{penalize}.} $\left\Vert\mathbf{V}\right\Vert_*-\left\Vert\mathbf{V}\right\Vert_2=0$.
Denote the objective function of problem \eqref{penalize} as $g\left(\mathbf{V};\rho\right)$. Since $\mathbf{V}_s$ is the optimal solution of problem \eqref{penalize} for penalty factor $\rho_s$, we have $g\left(\mathbf{V}_s;\rho_s\right)\le g\left(\mathbf{V}^\star;\rho_s\right)$, which leads to the inequality
\begin{equation}\label{eq43}
\begin{split}
&\relphantom{\le}f(\mathbf{V}_s)+\frac{1}{2\rho_s}\left(\left\Vert\mathbf{V}_s\right\Vert_*-\left\Vert\mathbf{V}_s\right\Vert_2\right)\\
&\le
f(\mathbf{V}^\star)+\frac{1}{2\rho_s}\left(\left\Vert\mathbf{V}^\star\right\Vert_*-\left\Vert\mathbf{V}^\star\right\Vert_2\right)=f(\mathbf{V}^\star),
\end{split}
\end{equation}
where the last equality holds as $\mathbf{V}^\star$ is the optimal solution of problem \eqref{eq23} and therefore the rank-one constraint must be satisfied. By rewriting \eqref{eq43}, we obtain
\begin{equation}\label{eq44}
\left\Vert\mathbf{V}_s\right\Vert_*-\left\Vert\mathbf{V}_s\right\Vert_2\le2\rho_s\left[f(\mathbf{V}^\star)-f(\mathbf{V}_s)\right].
\end{equation}
Suppose $\bar{\mathbf{V}}$ is a limit point of sequence $\{\mathbf{V}_s\}$ and there is an infinite subsequence $\mathcal{S}$ such that $\lim\limits_{s\in\mathcal{S}}\,\mathbf{V}_s=\bar{\mathbf{V}}$. By taking the limit as $s\to \infty$, $s\in\mathcal{S}$, on both sides of \eqref{eq44}, we have
\begin{equation}
\begin{split}
\left\Vert\bar{\mathbf{V}}\right\Vert_*-\left\Vert\bar{\mathbf{V}}\right\Vert_2&=\lim\limits_{s\in\mathcal{S}}\,\left(\left\Vert\mathbf{V}_s\right\Vert_*-\left\Vert\mathbf{V}_s\right\Vert_2\right)\\
&\le\lim\limits_{s\in\mathcal{S}}\,2\rho_s\left[f(\mathbf{V}^\star)-f(\mathbf{V}_s)\right]\overset{\rho_s\to0}{=}0,
\end{split}
\end{equation}
where the first equality holds due to the continuity of the function $\left\Vert\mathbf{X}\right\Vert_*-\left\Vert\mathbf{X}\right\Vert_2$.
Hence, we have $\left\Vert\bar{\mathbf{V}}\right\Vert_*-\left\Vert\bar{\mathbf{V}}\right\Vert_2=0$, so $\bar{\mathbf{V}}$ is feasible for problem \eqref{eq23}. Furthermore, by taking the limit as $s\to\infty$ for $s\in\mathcal{S}$ in \eqref{eq43}, we have by the non-negativity  of $\rho_s$ and of the term $\left\Vert\mathbf{V}_s\right\Vert_*-\left\Vert\mathbf{V}_s\right\Vert_2$ that
\begin{equation}
f\left(\bar{\mathbf{V}}\right)\le f\left(\bar{\mathbf{V}}\right)+\lim\limits_{s\in\mathcal{S}}\,\frac{1}{2\rho_s}\left(\left\Vert\mathbf{V}_s\right\Vert_*-\left\Vert\mathbf{V}_s\right\Vert_2\right)\le f\left(\mathbf{V}^\star\right).
\end{equation}
As $\bar{\mathbf{V}}$ is a feasible point whose objective value is no larger than that of the optimal solution $\mathbf{V}^\star$, we conclude that $\bar{\mathbf{V}}$, too, is an optimal solution for \eqref{eq23}, which completes the proof.

\bibliographystyle{IEEEtran}
\bibliography{bare_jrnl}

\end{document}